\documentclass[numberedappendix]{emulateapj}

\shorttitle{Mixed Frame Radiation Hydrodynamics}
\shortauthors{Krumholz, McKee, \& Klein}

\newcommand{\ltsim}{\protect\raisebox{-0.5ex}{$\:\stackrel{\textstyle <}
        {\sim}\:$}}
\newcommand{\gtsim}{\protect\raisebox{-0.5ex}{$\:\stackrel{\textstyle >}
        {\sim}\:$}}

\newcommand{\calp}{\mathcal{P}}
\newcommand{\kr}{\kappa_{\rm 0R}}
\newcommand{\kp}{\kappa_{\rm 0P}}
\newcommand{\ke}{\kappa_{\rm 0E}}
\newcommand{\kf}{\kappa_{\rm 0F}}
\newcommand{\kz}{\kappa_0}
\newcommand{\lp}{\lambda_{\rm P}}
\newcommand{\vecq}{\mathbf{q}}
\newcommand{\vecv}{\mathbf{v}}
\newcommand{\vecf}{\mathbf{f}}
\newcommand{\vecn}{\mathbf{n}}
\newcommand{\vecr}{\mathbf{r}}
\newcommand{\vecG}{\mathbf{G}}
\newcommand{\vecF}{\mathbf{F}}

\newcommand{\msun}{M_{\odot}}
\newcommand{\lsun}{L_{\odot}}
\newcommand{\taubf}{\mbox{\boldmath$\tau$}}
\newcommand{\betabf}{\mbox{\boldmath$\beta$}}
\newcommand{\veca}{\mathbf{a}}
\newcommand{\vecb}{\mathbf{b}}
\newcommand{\calA}{\mathcal{A}}
\newcommand{\calB}{\mathcal{B}}
\newcommand{\calI}{\mathcal{I}}

\begin{document}

\title{Equations and Algorithms for Mixed Frame\\Flux-Limited Diffusion
Radiation Hydrodynamics}

\slugcomment{Accepted for publication in the Astrophysical Journal Supplement, May 24, 2007}

\author{Mark R. Krumholz\footnote{Hubble Fellow}}
\affil{Department of Astrophysical Sciences, Princeton University,
Princeton, NJ 08544}
\email{krumholz@astro.princeton.edu}

\author{Richard I. Klein}
\affil{Department of Astronomy, University of California, Berkeley,
Berkeley, CA 94720, and Lawrence Livermore National Laboratory,
P.O. Box 808, L-23, Livermore, CA 94550}
\email{klein@astron.berkeley.edu}

\author{Christopher F. McKee}
\affil{Departments of Physics and Astronomy, University of California,
Berkeley, Berkeley, CA 94720}
\email{cmckee@astron.berkeley.edu}

\author{John Bolstad}
\affil{Lawrence Livermore National Laboratory, P.O. Box 808, L-23,
Livermore, CA 94550}
\email{bolstad@llnl.gov}

\begin{abstract}
We analyze the mixed frame equations of radiation hydrodynamics under
the approximations of flux-limited diffusion and a thermal radiation
field, and derive the minimal set of evolution equations that includes
all terms that are of leading order in any regime of non-relativistic
radiation hydrodynamics. Our equations are accurate to first order 
in $v/c$ in the static diffusion regime. In contrast, we show that
previous lower order derivations of these equations omit leading terms
in at least some regimes. In comparison to comoving frame formulations
of radiation hydrodynamics, our equations have the advantage that 
they manifestly conserve total energy, making them very well-suited to
numerical simulations, particularly with adaptive meshes. For systems
in the static diffusion regime, our analysis also suggests an
algorithm that is both simpler and faster than earlier comoving frame
methods. We implement this algorithm in the Orion adaptive mesh
refinement code, and show that it performs well in a range of
test problems.
\end{abstract}


\section{Introduction}

Astrophysical systems described by radiation hydrodynamics span a
tremendous range of scales and parameter regimes, from the interiors
of stars \citep[e.g.][]{kippenhahn94} to accretion disks around
compact objects \citep[e.g.][]{turner03} to dusty accretion flows
around massive protostars \citep[e.g.][]{krumholz05a, krumholz07a} to
galactic-scale flows onto AGN \citep[e.g.][]{thompson05}. All of these
systems have in common that matter and radiation are strongly
interacting, and that the energy and momentum carried by the
radiation field is significant in comparison to that carried by the
gas. Thus an accurate treatment of the problem must include
analysis of both the matter and the radiation, and of their
interaction.

Numerical methods exist to simulate such systems in a variety of
dimensionalities and levels of approximation. In three dimensions,
treatments of the matter and radiation fields generally adopt the
flux-limited diffusion approximation, first introduced by
\citet{alme73}, for reasons
of computational cost and simplicity \citep[e.g.][]{hayes06}. 
Flux-limited diffusion is optimal for treating continuum transfer in
a system such as an accretion disk, stellar atmosphere, or opaque
interstellar gas cloud where the majority of the interesting behavior
occurs in optically thick regions that are well described by pure
radiation diffusion, but there is a surface of optical depth unity
from which energy is radiated away. Applying pure diffusion to these
problems would lead to unphysically fast radiation from this surface,
so flux-limited diffusion provides a compromise that yields a
computationally simple and accurate description of the interior, while
also giving a reasonably accurate loss rate from the surface
\citep{castor04}.

However, the level of accuracy provided by this approximation has been
unclear because the equations of radiation hydrodynamics for
flux-limited diffusion have previously only been analyzed
to zeroth order in $v/c$. In contrast, several authors have
analyzed the radiation hydrodynamic equations in the general case
to beyond first order in $v/c$ \citep[e.g.][and references
therein]{mihalas99, castor04}. In a zeroth order treatment, one
neglects differences between quantities in the laboratory frame and
the comoving frame. The problem with this approach is that in an
optically thick fluid, the radiation flux only follows Fick's law
($\vecF\propto -\nabla E$) in the comoving frame, and in other frames
there is an added advective flux of radiation enthalpy, as first
demonstrated by \citet{castor72}. In
certain regimes (i.e.\ the dynamic diffusion limit -- see below)
this advective flux can dominate the diffusive flux \citep{mihalas01,
castor04}.

\citet{pomraning83} does give a flux-limiter usable to first order in
$v/c$, which is an approach to the problem of
flux-limiting with relativistic corrections that is an alternative to the one we
pursue in this paper. However, this approach
does not correctly handle the dynamic diffusion limit, a case that as
we show requires special attention because order $v^2/c^2$ terms can
be important. Furthermore, \citeauthor{pomraning83} derives his
flux-limiter directly from the transfer equation, so the computation
provides little insight into the relative importance
of radiation hydrodynamic terms, and the level of accuracy
obtained by using the uncorrected flux-limiter, the most common
procedure in astrophysical applications.

\citet{mihalas82} were the first to derive the
mixed frame equations of radiation hydrodynamics dynamics to order
$v/c$ in frequency-integrated and frequency-dependent forms, and gave
numerical algorithms for solving them.
\citet{lowrie99}, \citet{lowrie01}, and \citet{hubeny06} give alternate
forms of these equations, as well as numerical algorithms for solving
them. However,
these treatments require that one solve the radiation momentum
equation (and for the frequency-dependent equations calculate over many
frequencies as well), rather than adopt the flux-limited diffusion
approximation. While this preferable from a standpoint of accuracy,
since it allows explicit conservation of both momentum and energy and
captures the angular-dependence of the radiation field in a way that
diffusion methods cannot, treating the radiation momentum equation is
significantly more computationally costly than using flux-limited
diffusion, making it difficult to use in three-dimensional
calculations.

In this Paper we analyze the equations of radiation hydrodynamics
under the approximations that the radiation field has a thermal
spectrum and obeys the flux-limited diffusion approximation, and that
scattering is negligible for the system. Our goal is to derive an
accurate set of mixed frame equations, meaning that radiation
quantities are written in the lab frame, but fluid quantities, in
particular fluid opacities, are evaluated in the frame comoving with
the fluid. This formulation is optimal for three-dimensional
simulations, because writing radiation quantities in the lab frame
lets us use an Eulerian grid on which the radiative transfer problem
may be solved by any number of standard methods, while avoiding the
need to model the direction- and velocity-dependence of the lab frame
opacity and emissivity of a moving fluid.

In \S~\ref{equations} we begin from the general lab frame equations of
hydrodynamics to first order in
$v/c$, apply the flux-limited diffusion approximation in the frame
comoving with the gas where it is applicable, and transform the
appropriate radiation quantities into the lab frame, thereby deriving
the corresponding mixed frame equations suitable for implementation in
numerical simulations. We retain enough terms to ensure that we
achieve order unity accuracy in all regimes, and order $v/c$ accuracy
for static diffusion problems.
In \S~\ref{termanalysis} we assess the significance of the higher
order terms that appear in our equations, and consider where
treatments omitting them are acceptable, and where they are likely to
fail. We show that, in at least some regimes, the zeroth order
treatments most often used are likely to produce results that are
incorrect at order unity. We also compare our equations to the
comoving frame equations commonly used in other codes.
In \S~\ref{algorithm} we take advantage of the
ordering of terms we derive for the static diffusion regime to
construct a radiation hydrodynamic simulation algorithm for static
diffusion problems that is
simpler and faster than those now in use, which we implement in the
Orion adaptive mesh refinement code. In \S~\ref{tests} we demonstrate it
in a selection of test problems. Finally, we summarize our results in
\S~\ref{summary}.

\section{Derivation of the Equations}
\label{equations}

In the discussion that follows, we adopt the convention of writing
quantities measured in the frame comoving with a fluid with a
subscript zero. Quantities in the lab frame are written without
subscripts. We write scalars in italics (e.g.\ $a$),
vectors in boldface (e.g.\ $\veca$), and rank two tensors in
calligraphy (e.g.\ $\calA$). We indicate tensor contractions over a
single index by dots (e.g.\ $\veca\cdot\vecb=a^i b_i$), tensor
contractions over two indices by colons (e.g.\
$\calA\colon\calB=A^{ij} B_{ij}$), and tensor products of
vectors without any operator symbol (e.g.\ $(\veca\vecb)^{ij}
=a^i b^j$).
Also note that we follow the standard convention in
radiation hydrodynamics rather than the standard in astrophysics, in
that when we refer to an opacity $\kappa$ we mean the total
opacity, measured in units of inverse length, rather than the
specific opacity, measured in units of length squared divided by
mass. Since we are neglecting scattering, we may set the
extinction $\chi=\kappa$.

\subsection{Regimes of Radiation Hydrodynamics}

Before beginning our analysis, it is helpful to examine some
characteristic dimensionless numbers for a radiation hydrodynamic
system, since evaluating these quantities provides a
useful guide to how we should analyze our equations. 
Let $\ell$ be the characteristic size of the system under
consideration, $u$ be the characteristic velocity in this system, and
$\lp\sim 1/\kappa$, be the photon mean free path.
Following \citet{mihalas99}, we can define three distinct
limiting cases by considering the dimensionless ratios $\tau\equiv
\ell/\lp$, which characterizes the optical depth of the system, and
$\beta \equiv u/c$, which characterizes how relativistic it is. Since we focus on non-relativistic systems, we assume $\beta\ll 1$. We
term the case $\tau \ll 1$, in which the radiation and gas are weakly
coupled, the \textit{streaming} limit. If $\tau \gg 1$ then radiation
and gas are strongly coupled, and the system is in the diffusion
limit. We can further subdivide the diffusion limit into the cases
$\beta \gg \tau^{-1}$ and $\beta \ll \tau^{-1}$. The former is the
\textit{dynamic diffusion} limit, while the latter is the
\textit{static diffusion} limit. In summary, the limiting cases are
\begin{eqnarray}
\tau\ll 1, & & \mbox{ (streaming limit)} \\
\tau\gg 1, & \;\beta \tau \ll 1, & \mbox{ (static diffusion limit)} \\
\tau\gg 1, & \;\beta \tau \gg 1, & \mbox{ (dynamic diffusion limit)}.
\end{eqnarray}

Physically, the distinction between static and dynamic diffusion
is that in dynamic diffusion radiation is principally
transported by advection by gas, so that terms describing the work
done by radiation on gas and the advection of radiation enthalpy
dominate over terms describing either diffusion or emission and
absorption. In the static diffusion limit the opposite holds. 
A paradigmatic example of a dynamic diffusion system is a stellar
interior. The optical depth from the core to the surface of
the Sun is $\tau \sim 10^{11}$, and typical convective and rotational
velocities are $\gg 10^{-11} c = 0.3$ cm s$^{-1}$, so the Sun is
strongly in the dynamic diffusion regime. In contrast, an example of a
system in the static diffusion limit is a relatively cool, dusty,
outer accretion disk around a forming massive protostar, as studied
e.g.\ by \citet{krumholz07a}. The specific opacity of gas with the standard
interstellar dust abundance to infrared photons is $\kappa/\rho
\sim 1$ cm$^2$ g$^{-1}$, and at distances of more than a few AU
from the central star the density is generally $\rho \ltsim 10^{-12}$ g
cm$^{-3}$. For a disk of scale height $h\sim 10$ AU, the optical depth
to escape is
\begin{eqnarray}
\tau^{-1} & \approx &
6.7\times 10^{-3}
\left(\frac{\kappa/\rho}{\mbox{cm$^2$ g$^{-1}$}}\right)^{-1}
\nonumber \\
& & \qquad \left(\frac{\rho}{10^{-12}\mbox{ g cm$^{-3}$}}\right)^{-1}
\left(\frac{h}{10\mbox{ AU}}\right)^{-1}.
\end{eqnarray}
The velocity is roughly the Keplerian speed, so
\begin{equation}
\beta \approx 1.4 \times 10^{-4} 
\left(\frac{M_*}{10\;\msun}\right)^{1/2}
\left(\frac{r}{10\mbox{ AU}}\right)^{-1/2},
\end{equation}
where $M_*$ is the mass of the star and $r$ is the distance from
it. Thus, this system is in a static diffusion regime by roughly two
orders of magnitude.

In the analysis that follows, our goal will be to obtain expressions
that are accurate for the leading terms in all regimes. This is
somewhat tricky, particularly for diffusion problems, because we are
attempting to expand our equations simultaneously in the two small
parameters $\beta$ and $1/\tau$. The most common approach in radiation
hydrodynamics is to expand expressions in powers of $\beta$ alone,
and only analyze the equations in terms of $\tau$ after dropping
terms of high order in $\beta$. However, this approach can produce
significant errors, because terms in the radiation hydrodynamic
equations proportional to the opacity are multiplied by a quantity of
order $\tau$. Thus, in our derivation we will repeatedly encounter
expressions proportional to $\beta^2\tau$, and in a problem that is
either in the dynamic diffusion limit or close to it ($\beta\tau
\gtsim 1$), it is inconsistent to drop these terms while retaining
ones that are of order $\beta$. We therefore retain all terms up to
order $\beta^2$ in our derivation unless we explicitly check that
they are not multiplied by terms of order $\tau$, and can therefore be
dropped safely.

\subsection{The Equations of radiation hydrodynamics}
\label{equationderivation}

We now start our derivation, beginning from the lab frame equations of
radiation hydrodynamics \citep{mihalas82, mihalas99, mihalas01}
\begin{eqnarray}
\label{massconservation}
\frac{\partial \rho}{\partial t} + \nabla\cdot (\rho \vecv) & = & 0 \\
\label{mom1}
\frac{\partial}{\partial t} (\rho \vecv) + \nabla \cdot (\rho \vecv \vecv) 
& = & -\nabla P + \vecG \\
\label{en1}
\frac{\partial}{\partial t} (\rho e) + \nabla \cdot \left[\left(\rho e +
P\right) \vecv\right] & = & c G^0 \\
\label{raden1}
\frac{\partial E}{\partial t} + \nabla \cdot \vecF &=& - c G^0 \\
\label{radmom1}
\frac{1}{c^2} \frac{\partial \vecF}{\partial t} + \nabla\cdot\calp & =
& -\vecG
\end{eqnarray}
where $\rho$, $\vecv$, $e$, and $P$ are the density, velocity,
specific energy (thermal plus kinetic), and thermal pressure of the
gas, $E$, $\vecF$, and $\calp$ are the radiation energy density, flux,
and pressure tensor,
\begin{eqnarray}
c E & = & \int_0^{\infty} d\nu \int d\Omega \, I(\vecn, \nu) \\
\vecF & = & \int_0^{\infty} d\nu \int d\Omega \, \vecn I(\vecn, \nu) \\
c \calp & = & \int_0^{\infty} d\nu \int d\Omega \, \vecn \vecn I(\vecn,
\nu),
\end{eqnarray}
$(G^0,\vecG)$ is the radiation four-force density
\begin{eqnarray}
cG^0 & = & \int_0^{\infty} d\nu \int d\Omega\,[\kappa(\vecn,\nu)
I(\vecn, \nu) - \eta(\vecn,\nu)], \\
c\vecG & = & \int_0^{\infty} d\nu \int d\Omega\,
[\kappa(\vecn,\nu) I(\vecn, \nu) - \eta(\vecn,\nu)] \vecn,
\end{eqnarray}
and $I(\vecn,\nu)$ is the intensity of the radiation field at
frequency $\nu$ traveling in direction $\vecn$.
Here $\kappa(\vecn,\nu)$ and $\eta(\vecn,\nu)$ are the direction- and
frequency-dependent radiation absorption and emission coefficients in
the lab frame. Intuitively, we can understand $cG^0$ as the rate of
energy absorption from the radiation field minus the rate of energy
emission for the fluid, and $\vecG$ as the rate of momentum absorption
from the radiation field minus the rate of momentum
emission. Equations (\ref{massconservation}) -- (\ref{en1}) are
accurate to first order in $v/c$, while equations (\ref{raden1}) --
(\ref{radmom1}) are exact. Note that no terms involving opacity or
optical depth appear explicitly in any of these equations, so the fact
that they are accurate to first order in $\beta$ means that they
include all the leading order terms.

In order to derive the mixed-frame equations, we must now evaluate the radiation four-force $(G^0,\vecG)$ in terms of lab frame radiation quantities and comoving frame emission and absorption coefficients. \citet{mihalas01} show that, if the flux spectrum of the radiation is
direction-independent, the radiation four-force on a
thermally-emitting material to all orders in $v/c$ is given in terms
of moments of the radiation field by
\begin{eqnarray}
G^0 & = & \gamma [\gamma^2 \ke + (1-\gamma^2) \kf] E
- \gamma\kp a_R T_0^4 
\nonumber \\
& & {} - \gamma (\vecv\cdot\vecF/c^2)
[\kf - 2\gamma^2 (\kf-\ke)]
\nonumber \\
& & {} 
- \gamma^3 (\kf-\ke) (\vecv\vecv)\colon\calp/c^2, \\
\vecG & = & \gamma\kf (\vecF/c) 
- \gamma \kp a_R T_0^4 (\vecv/c)
\nonumber \\
& & {} - 
[\gamma^3 (\kf - \ke) (\vecv/c) E + \gamma\kf
(\vecv/c)\cdot \calp]
\nonumber \\
& & {} + \gamma^3 (\kf-\ke) [ 2 \vecv\cdot\vecF/c^3 -
(\vecv\vecv)\colon\calp/c^3] \vecv,
\end{eqnarray}
where $\gamma = 1/\sqrt{1-v^2/c^2}$ is the Lorentz
factor and $T_0$ is the gas temperature. The three
opacities that appear are the Planck-, energy-, and flux-mean
opacities, which are defined by
\begin{eqnarray}
\kp & = & 
\frac{\int_0^{\infty} d\nu_0 \, \kappa_0(\nu_0) 
B(\nu_0, T_0)}{B(T_0)} \\
\ke & = &
\frac{\int_0^{\infty} d\nu_0 \, \kappa_0(\nu_0) 
E_0(\nu_0)}{E_0} \\
\label{fluxmeandef}
\kf & = &
\frac{\int_0^{\infty} d\nu_0 \, \kappa_0(\nu_0) 
\vecF_0(\nu_0)}{\vecF_0},
\end{eqnarray}
where $E_0(\nu_0)$ and $\vecF_0(\nu_0)$ are the comoving frame 
radiation energy and flux per
unit frequency, $E_0$ and $\vecF_0$ are the corresponding
frequency-integrated energy and flux,
and $B(\nu,T)= (2 h \nu^3/c^2) / (e^{h\nu/k_B T} - 1)$
and $B(T) = c a_R T^4/(4\pi)$ are the frequency-dependent and
frequency-integrated Planck functions. 

Note that we have implicitly
assumed that the opacity and emissivity are directionally-independent
in the fluid rest frame, which is the case for any conventional material. We have
also assumed that the flux spectrum is independent of direction,
allowing us to replace the flux-mean opacity vector with a
scalar. This may not be the case for an optically thin system, or one
in which line transport is important, but since we are limiting our
application to systems to which we can reasonably apply the diffusion
approximation, this is not a major limitation.

To simplify $(G^0, \vecG)$, first we assume that the radiation has a
blackbody spectrum, so that $E_0(\nu_0)\propto B(\nu_0, T_0)$. In this
case, clearly 
\begin{equation}
\ke = \kp.
\end{equation}
Second, we adopt the
flux-limited diffusion approximation (see below), so in optically
thick parts of the flow $\vecF_0(\nu_0) \propto
-\nabla E_0(\nu_0)/\kappa_0(\nu_0)$ (Fick's Law). This implies that
$\vecF_0(\nu_0) \propto -[\partial B(\nu_0, T_0)/\partial T_0] (\nabla
T_0) / \kappa_0(\nu_0)$, and substituting this into (\ref{fluxmeandef})
shows that the flux-mean opacity $\kf$ is equal to the
Rosseland-mean opacity, defined by
\begin{equation}
\kr^{-1} = \frac{\int_0^{\infty} d\nu_0 \,
\kappa_0(\nu_0)^{-1} 
\frac{\partial B(\nu_0,T_0)}{\partial T_0}}
{\int_0^{\infty} d\nu_0 \,
\frac{\partial B(\nu_0,T_0)}{\partial T_0}}.
\end{equation}
In optically thin parts of the flow, $|\vecF_0(\nu_0)|\rightarrow c
E_0(\nu_0)$, so in principle we should have
$\kf=\ke$. However, interpolating between these cases
is complex, and the flux-limited diffusion approximation is of
limited accuracy for optically thin flows in complex
geometries. Moreover, our approximation that the radiation spectrum is
that of a blackbody at the local radiation temperature is itself
problematic in the optically thin limit, so setting
$\kf=\kp$ would not necessarily be more accurate than
using $\kr$. We therefore choose to
optimize our accuracy in the optically thick part of the flow 
and set 
\begin{equation}
\kf=\kr.
\end{equation}
With these two approximations, the only two opacities remaining in
our equations are $\kr$ and $\kp$, both of which are
independent of the spectrum of the radiation field and the direction
of radiation propagation, and which may therefore be
tabulated as a function of temperature for a given material once and
for all.

Next, we expand $(G^0,\vecG)$ in powers of $v/c$, retaining terms to
order $v^2/c^2$. In performing this
expansion, we note that $|\vecF|\le cE$, and $\mbox{Tr}(\calp)=E$. The
resulting expression for the radiation four-force is
\begin{eqnarray}
G^0 & = & \kp \left(E - \frac{4\pi B}{c}\right) 
+ \left(\kr-2\kp\right) \frac{\vecv\cdot\vecF}{c^2}
\nonumber \\
& & {}
+ \frac{1}{2} \left(\frac{v}{c}\right)^2 \left[2 (\kp-\kr) E + \kp \left(E-\frac{4\pi B}{c}\right)\right]
\nonumber \\
& & {}
+ (\kp-\kr) \frac{\vecv\vecv}{c^2}\colon\calp + O\left(\frac{v^3}{c^3}\right)
\label{G01}
\\
\vecG & = & \kr \frac{\vecF}{c} 
+ \kp \left(\frac{\vecv}{c}\right) \left(E - \frac{4\pi
B}{c}\right)
\nonumber \\
& & {}
- \kr\left[\frac{\vecv}{c}E + \frac{\vecv}{c}\cdot \calp\right]
+ \frac{1}{2}\left(\frac{v}{c}\right)^2 \kr\frac{\vecF}{c}
\nonumber \\
& & {}
+ 2 (\kr-\kp) \frac{(\vecv\cdot\vecF)\vecv}{c^3} +
O\left(\frac{v^3}{c^3}\right)
\label{Gvec1}
\end{eqnarray}

It is helpful at this point, before we making any further
approximations, to examine the scalings of these terms with the help
of our dimensionless parameters $\beta$ and $\tau$. In the streaming
limit, radiation travels freely at $c$ and emission and
absorption of radiation by matter need not balance, so $|\vecF|\sim c
E$ and $4\pi B/c-E \sim E$. For static diffusion, \citet{mihalas99}
show that $|\vecF|\sim c E/\tau$ and $4\pi B/c - E \sim E/\tau^2$. For
dynamic diffusion, radiation travels primarily by advection, so
$|\vecF| \sim v E$. We show in Appendix \ref{dyndiffusionscaling} that
for dynamic diffusion $4\pi B/c - E \sim \beta^2 E$. Note that the
scaling $4\pi B/c-E\sim (\beta/\tau) E$ given in \citet{mihalas99}
appears to be incorrect, as we show in the Appendix. Using these
values, we obtain the scalings shown in Table \ref{Gscalings} for the
terms in (\ref{G01}) and (\ref{Gvec1}).

\begin{deluxetable*}{ccccc}
\tablecaption{Scalings of terms in the radiation four-force density
\label{Gscalings}}
\tablehead{
\colhead{$G^0$ or $\vecG$} &
\colhead{Term} &
\colhead{Streaming} & 
\colhead{Static Diffusion} & 
\colhead{Dynamic Diffusion}
}
\startdata
$G^0$ & $\kp(E-4\pi B/c)$ & $\taubf$ & $\mathbf{1/\taubf}$ & $\mathbf{\betabf^2\taubf}$ \\
$G^0$ & $(\kr-2\kp) (\vecv\cdot\vecF/c^2)$ & $\beta \tau$
& $\beta$ & $\mathbf{\betabf^2 \taubf}$ \\
$G^0$ & $(v/c)^2(\kp-\kr) E$ & $\beta^2 \tau$ & $\beta^2 \tau$ & $\mathbf{\betabf^2 \taubf}$ \\
$G^0$ & $(1/2)(v/c)^2 \kp (4\pi B/c-E)$ & $\beta^2 \tau$ & $\beta^2/\tau$ & $\beta^4 \tau$ \\
$G^0$ & $(\kr-\kp) (\vecv\vecv/c^2)\colon\calp$ & $\beta^2 \tau$ & $\beta^2 \tau$ & $\mathbf{\betabf^2 \taubf}$ \\
$\vecG$ & $\kr\vecF/c$ & $\mathbf{\taubf}$ & $\mathbf{1}$ & $\betabf \taubf$ \\
$\vecG$ & $\kp(\vecv/c)(4\pi B/c-E)$ & $\beta \tau$ & $\beta/\tau$ & $\beta^3 \tau$ \\
$\vecG$ & $\kr[(\vecv/c)E + (\vecv/c)\cdot\calp]$ & $\beta \tau$ & $\beta \tau$ & $\betabf\taubf$ \\
$\vecG$ & $(1/2)(v/c)^2 \kr \vecF/c$ & $\beta^2 \tau$ & $\beta^2$ & $\beta^3 \tau$ \\
$\vecG$ & $2 (\kr-\kp) (\vecv\cdot\vecF)\vecv/c^3$ & $\beta^2 \tau$ & $\beta^2$ & $\beta^3 \tau$ \\
\enddata
\tablecomments{
Col. (1): Whether the term appears in $G^0$ or $\vecG$.
Col. (3)-(5): All scalings are normalized to $E/\ell$. The scalings that are of leading order in each regime are boldfaced.
}
\end{deluxetable*}

The Table shows that, despite the fact that we have kept all terms
that are formally order $\beta^2$ or more, in fact we only have
leading-order accuracy in the dynamic diffusion limit, because in this
limit the order unity and order $\beta$ terms in $G^0$ vanish to order
$\beta^2$. To obtain the next-order terms, we would have had to write
$G^0$ to order $\beta^3$. A corollary of this is that
treatments of the dynamic diffusion limit that do not retain order
$\beta^2$ terms are likely to produce equations that are
incorrect at order unity, since they will have dropped terms that are
of the same order as the ones that have been retained.

At this point we could begin dropping terms that are insignificant at
the order to which we are working, but it is cumbersome to construct a
table analogous to Table \ref{Gscalings} at every step of our
derivation. It is more convenient to continue our analysis retaining
all the terms in (\ref{G01}) and (\ref{Gvec1}), and to drop terms only
periodically.

Before moving on, there is a subtlety in (\ref{G01}) and (\ref{Gvec1})
that is worth commenting on. Consider a gray fluid, one in which
$\kr=\kp=\kz$. In $cG^0$, the term that describes the work done by
radiation, $-\kz \vecv \cdot \vecF/c$, has the opposite sign from what
one might naively expect. Using $cG^0$ in the gas energy equation
(\ref{en1}) in this case implies that the gas energy increases when
$\vecv$ and $\vecF$ are anti-aligned, i.e.\ when gas moves into an
oncoming photon flux. We can understand the origin of this somewhat
counter-intuitive behavior by considering the example of a fluid in
thermal equilibrium with a radiation field in its rest frame (i.e.\
$4\pi B = c E_0$). In the comoving frame, the
radiation four-force behaves as one intuitively expects: at leading
order the rate at which the radiation field transfers momentum density
to the gas is $\vecG_0 = \kz \vecF_0/c$, and the rate at which the
gas energy density changes as a result is $c G^0_0 = \kz \vecv\cdot
\vecF_0/c$ \citep[their equations 53a and 53b]{mihalas01}. Thus, gas
loses energy when it moves opposite the direction of the flux, and hence opposite the force. 

However, now consider the fluid as seen by an observer in a frame
boosted by velocity $-\vecv$ relative to the fluid.
The observer sees the radiation energy density as $E$, which differs
from $E_0$ by $2 \vecv \cdot \vecF/c^2$ (see
equation \ref{etransform}), and this difference is the reason that the
work term in $G^0$ is $-\kz \vecv\cdot\vecF/c^2$.
Physically, this happens because an observer who sees the fluid moving
at velocity $\vecv$ also sees the radiation and gas as being out of
thermal equilibrium ($4\pi B \neq c E$), since $E$ and $E_0$ are
different. This disequilibrium leads the radiation and gas to exchange
energy at a rate that is opposite in direction and twice as large as
the radiation work, $\kz\vecv\cdot\vecF/c$. This is why the ``work"
term has the opposite sign than the one we might expect. Thus, for the
rest of this paper, while for convenience we continue to refer to
$(\kr-2\kp) \vecv \cdot \vecF/c$ and the terms to which it gives rise
as ``work" terms, it is important to keep in mind that in reality this
term contains contributions from two different effects of comparable
magnitude, the ``Newtonian work" $\kr \vecv\cdot\vecF/c$ and the
post-Newtonian term $-2\kp \vecv\cdot\vecF/c$ describing the imbalance
between emission and absorption that an observer sees solely because
the fluid is moving.

With this point understood, we now adopt the flux-limited diffusion
approximation \citep{alme73}, under which we drop the radiation momentum
equation (\ref{radmom1}) and set the radiation flux in the
comoving frame to
\begin{equation}
\label{fldflux}
\vecF_0 = -\frac{c\lambda}{\kr} \nabla E_0,
\end{equation}
where $\lambda$ is a dimensionless number called the
flux-limiter. Many functional forms for $\lambda$ are possible. For
the code implementation we describe later, we adopt the
\citet{levermore81} flux-limiter, given by
\begin{eqnarray}
\lambda & = & \frac{1}{R} \left(\coth R - \frac{1}{R}\right) \\
R & = & \frac{|\nabla E_0|}{\kr E_0}.
\end{eqnarray}
However, our derivation is independent of this choice. Regardless of
their exact functional form, all flux limiters have the property that in
an optically thick medium $\lambda\rightarrow 1/3$, thereby giving
$\vecF_0\rightarrow -[c/(3\kr)]\nabla E_0$, the
correct value for diffusion. In an optically thin medium,
$\lambda\rightarrow (\kr E_0/|\nabla E_0|) \vecn_0$, where $\vecn_0$ is the
unit vector antiparallel to $\nabla E_0$, so the flux
approaches $\vecF_0\rightarrow c E_0 \vecn_0$, and the propagation
speed of radiation is correctly limited to $c$.

For the \citeauthor{levermore81} flux-limiter we adopt, 
the corresponding approximate value for the
radiation pressure tensor is \citep{levermore84}
\begin{equation}
\label{fldpressure}
\calp_0 = \frac{E_0}{2} \left[(1-R_2)\calI + (3 R_2 - 1)
\vecn_0 \vecn_0\right],
\end{equation}
where $\calI$ is the identity tensor of rank 2 and
\begin{equation}
R_2 = \lambda + \lambda^2 R^2.
\end{equation}
Physically, this approximation interpolates between the behavior in
very optically thick regions, where $R_2\rightarrow 1/3 +
O(1/\tau^2)$, the radiation pressure is isotropic, and off-diagonal
components vanish, and optically thin regions, where $R_2 \rightarrow
1$ and the radiation pressure tensor is zero orthogonal to $\vecn_0$ and
$E_0$ parallel to it.

Note that for pure diffusion \citet{mihalas99} and \citet{castor04}
show that the pressure tensor reduces to $(E_0/3) \calI$ plus
off-diagonal elements of order $\beta/\tau$ or $\beta^2$. Our
approximation does not quite reproduce this, since in the
diffusion limit it gives $\calp_0 = (E_0/3) \calI$ plus
off-diagonal elements of order $\tau^{-2}$. We might therefore worry
that, in the static diffusion regime where $\beta \ll \tau^{-1}$, we
will have an incorrect term. However, examination of our final
equations below shows that all terms arising from off-diagonal
elements of $\calp_0$ are smaller than order $\beta$ in the static
diffusion limit, so adopting the \citet{levermore84} approximation for
the pressure tensor does
not introduce any incorrect terms at order $\beta$ in the final
equations.

To use the approximations (\ref{fldflux}) and (\ref{fldpressure}) to
evaluate the radiation four-force, we must Lorentz transform them to
express the radiation quantities in the lab frame. The Lorentz
transforms for the energy, flux, and pressure
to second order in $v/c$ are \citep{mihalas99}
\begin{eqnarray}
E & = & E_0 + 2\frac{\vecv\cdot\vecF_0}{c^2} + \frac{1}{c^2}\left[v^2 E_0 + (\vecv\vecv)\colon\calp_0\right]
\label{etransform}
 \\
\vecF & = & \vecF_0 + \vecv E_0 + \vecv\cdot\calp_0 + \frac{1}{2c^2} \left[v^2 \vecF_0+3 \vecv(\vecv\cdot\vecF_0)\right]
\label{ftransform}
\\
\calp & = & \calp_0 + \frac{\vecv\vecF_0 + \vecF_0\vecv}{c^2} +
\frac{1}{c^2}\left[\vecv\vecv E_0 + \vecv(\vecv\cdot\calp_0)\right].
\label{ptransform}
\end{eqnarray}
Note that in the expression for $\calp$ we have simplified the final
term using the fact that $\calp_0$ is a symmetric tensor.

Using the same scaling arguments we used to construct Table
\ref{Gscalings}, we see that $\calp$ and $\calp_0$ differ at order
$\beta$ in the streaming limit, at order $\beta/\tau$ for static
diffusion, and at order $\beta^2$ for dynamic diffusion. Since this is
below our accuracy goal, we need not distinguish
$\calp$ and $\calp_0$. The same is true of $E$ and $E_0$. However, $\vecF$ is different. In the comoving frame in an optically thick system, one is in the static diffusion regime, so $\vecF_0 \sim c E_0/\tau$. Since $\vecv E_0$ and $\vecv\cdot\calp_0$ are of order $\beta c E_0$, and in dynamic diffusion $\beta \gg 1/\tau$, this means that $\vecv E_0$ and $\vecv\cdot\calp_0$ are the dominant components of $\vecF$ in dynamic diffusion, and must therefore be retained. Thus, 
\begin{equation}
\label{labflux}
\vecF = -\frac{c\lambda}{\kr} \nabla E + \vecv E +
\vecv\cdot\calp,
\label{fluxtransform}
\end{equation}
which is simply the rest frame flux plus terms describing the
advection of radiation enthalpy.

Substituting (\ref{fldpressure}) with $\calp=\calp_0$ and
(\ref{fluxtransform}) into the four-force density (\ref{G01}) and
(\ref{Gvec1}), and continuing to retain terms to order $v^2/c^2$,
gives
\begin{eqnarray}
G^0 & = & \kp \left(E-\frac{4\pi B}{c}\right)
+ \left(\frac{\lambda}{c}\right)\left(2\frac{\kp}{\kr}-1\right)
\vecv \cdot \nabla E
\nonumber \\
& & {}
- \frac{\kp}{c^2} E \left[\frac{3-R_2}{2} v^2 + \frac{3
R_2-1}{2} (\vecv\cdot\vecn)^2\right] 
\nonumber \\
& & {}
+ \frac{1}{2}\left(\frac{v}{c}\right)^2 \kp \left(E-\frac{4\pi
B}{c}\right)
\label{gztrans}
\\
\vecG & = & - \lambda \nabla E + \kp \frac{\vecv}{c} 
\left(E - \frac{4\pi B}{c}\right)
\nonumber \\
& & {}
-\frac{1}{2}\left(\frac{v}{c}\right)^2 \lambda \nabla E 
\nonumber \\
& & {}
+ 2\lambda\left(\frac{\kp}{\kr}-1\right) \frac{(\vecv\cdot\nabla
E)\vecv}{c^2}.
\label{gtrans}
\end{eqnarray}
Here $\vecn$ is the unit vector antiparallel to $\nabla E$.
We again remind the reader that, although these equations contain
terms of order $\beta^2$, they are not truly accurate to order
$\beta^2$ because we did not retain all the $\beta^2$ when applying the
Lorentz transform to the flux and pressure. However, these equations
include all the terms that appear at the order of accuracy to which we
are working, and by retaining terms of order $\beta^2$ we guarantee
that these terms will be preserved.

Inserting $(G^0,\vecG)$ and the lab frame flux (\ref{labflux}) into the gas momentum and energy
equations (\ref{mom1}) and (\ref{en1}), and the radiation energy equation
(\ref{raden1}), and again retaining terms to order $v^2/c^2$ gives
\begin{eqnarray}
\frac{\partial}{\partial t}(\rho \vecv) & = &
- \nabla\cdot(\rho\vecv\vecv) -\nabla P 
- \lambda \nabla E 
\nonumber \\
& & {}
 - \kp \frac{\vecv}{c^2} 
\left(4\pi B - c E\right)
-\frac{1}{2}\left(\frac{v}{c}\right)^2 \lambda \nabla E 
\nonumber \\
& & {}
+ 2\lambda\left(\frac{\kp}{\kr}-1\right) \frac{(\vecv\cdot\nabla E)\vecv}{c^2}.
\label{mom2}\\
\frac{\partial}{\partial t}(\rho e) & = &
- \nabla\cdot[(\rho e + P)\vecv] - \kp (4\pi B - c E) 
\nonumber \\
& & {}
+ \lambda\left(2\frac{\kp}{\kr}-1\right) \vecv\cdot\nabla E 
\nonumber \\
& & {} 
- \frac{\kp}{c} E \left[\frac{3 - R_2}{2} v^2 + 
\frac{3 R_2 - 1}{2} (\vecv\cdot\vecn)^2
\right] 
\nonumber \\
& & {}
- \frac{1}{2}\left(\frac{v}{c}\right)^2 \kp \left(4 \pi B - c E\right)
\label{en2}
\\
\frac{\partial}{\partial t}E
& = &
\nabla\cdot\left(\frac{c\lambda}{\kr} \nabla E\right) + \kp (4\pi B - c E)
\nonumber \\
& & {}
-\lambda \left(2\frac{\kp}{\kr}-1\right) \vecv \cdot \nabla E
\nonumber \\
& & {} 
+\frac{\kp}{c} E \left[
\frac{3-R_2}{2} v^2 + \frac{3R_2-1}{2} (\vecv\cdot\vecn)^2
\right]
\nonumber \\
& & {}
- \nabla \cdot \left[
\frac{3-R_2}{2} \vecv E + \frac{3 R_2-1}{2} \vecv\cdot(\vecn\vecn)
E\right]
\nonumber \\
& & {}
+ \frac{1}{2}\left(\frac{v}{c}\right)^2 \kp \left(4 \pi B - c E\right).
\label{raden2}
\end{eqnarray}
At this point we construct Table \ref{eqscalings} showing the scalings of the radiation terms to see which must be retained and which are superfluous. In constructing the table, we take spatial derivatives to be of characteristic scaling $1/\ell$, i.e.\ we assume that radiation quantities vary on a size scale of the system, rather than over a size scale of the photon mean free path. In the streaming limit, $\lambda\sim \tau$ and $R_2\sim 1+O(\tau)$. In the diffusion limit $\lambda\sim 1/3$ and $R_2\sim 1/3+O(\tau^{-2})$.

\begin{deluxetable*}{ccccc}
\tablecaption{Scalings of terms in the conservation equations
\label{eqscalings}}
\tablehead{
\colhead{Equation} &
\colhead{Term} &
\colhead{Streaming} & 
\colhead{Static Diffusion} & 
\colhead{Dynamic Diffusion}
}
\startdata
M & $\lambda\nabla E$ & $\taubf$ & $\mathbf{1}$ & $\mathbf{1}$ \\
M & $\kp (\vecv/c^2) (4\pi B - c E)$ & $\beta\tau$ & $\beta/\tau$ &
$\beta^3\tau$ \\
M & $(1/2)(v/c)^2 \lambda \nabla E$ & $\beta^2 \tau$ & $\beta^2$ & $\beta^2$ \\
M & $2\lambda(\kp/\kr-1) (\vecv\cdot\nabla E)\vecv/c^2$ & $\beta^2\tau$ & $\beta^2$ & $\beta^2$ \\
G and R & $\kp (4\pi B - c E)$ & $\taubf$ & $\mathbf{1/\taubf}$ 
& $\mathbf{\betabf^2 \taubf}$ \\
G and R & $\lambda(2\kp/\kr-1) \vecv\cdot\nabla E$ & $\beta \tau$ &
$\beta$ & $\betabf$ \\
G and R & $\kp(v^2/c) [(3-R_2)/2]E$ & $\beta^2 \tau$ &
$\beta^2 \tau$ & $\mathbf{\betabf^2 \taubf}$ \\
G and R & $\kp[(\vecv\cdot\vecn)^2/c] [(3R_2 - 1)/2]E$ &
$\beta^2 \tau$ & $\beta^2/\tau$ & $\beta^2/\tau$ \\
G and R & $(1/2)(v/c)^2 \kp (c E - 4\pi B)$ & $\beta^2\tau$ & $\beta^2/\tau$ & $\beta^4\tau$ \\
R & $\nabla \cdot[(c\lambda/\kr)\nabla E]$ & $\mathbf{1}$ 
& $\mathbf{1/\taubf}$ & $1/\tau$ \\ 
R & $\nabla \cdot\{[(3-R_2)/2] \vecv E\}$ & $\beta$ &
$\beta$ & $\betabf$ \\
R & $\nabla \cdot\{[(3 R_2-1)/2] \vecv\cdot(\vecn\vecn) E\}$ &
$\beta$ & $\beta/\tau^2$ & $\beta/\tau^2$ \\
\enddata
\tablecomments{
Col. (1): Which equation the term appears in. M = momentum (\ref{mom2}), G = gas energy (\ref{en2}), R = radiation energy (\ref{raden2}). Col. (3)-(5): All scalings are normalized to $E/l$ for the momentum equation, and $c E/\ell$ for the energy equations. Scalings that are of leading order in each regime for each equation are boldfaced.
}
\end{deluxetable*}

Using Table \ref{eqscalings} to drop all terms that are not significant at leading order in any regime, we arrive at our final equations:
\begin{eqnarray}
\frac{\partial}{\partial t}(\rho \vecv) & = &
- \nabla\cdot(\rho\vecv\vecv) -\nabla P 
- \lambda \nabla E 
\label{momentumconservation}\\
\frac{\partial}{\partial t}(\rho e) & = &
- \nabla\cdot[(\rho e + P)\vecv] - \kp (4\pi B - c E) 
\nonumber \\
& & {}
+ \lambda\left(2\frac{\kp}{\kr}-1\right) \vecv\cdot\nabla E 
\nonumber \\
& & {} 
- \frac{3-R_2}{2} \kp \frac{v^2}{c} E
\label{gasenergy}
\\
\frac{\partial}{\partial t}E
& = &
\nabla\cdot\left(\frac{c\lambda}{\kr} \nabla E\right) + \kp (4\pi B - c E)
\nonumber \\
& & {}
-\lambda \left(2\frac{\kp}{\kr}-1\right) \vecv \cdot \nabla E
\nonumber \\
& & {} 
+ \frac{3-R_2}{2} \kp \frac{v^2}{c} E
\nonumber \\
& & {} 
 - \nabla\cdot\left(\frac{3-R_2}{2}\vecv E\right).
\label{radenergy}
\end{eqnarray}
These represent the equations of momentum conservation for the gas,
energy conservation for the gas, and energy conservation for the
radiation field, which, together with the equation of mass
conservation (\ref{massconservation}), fully describe the system under
the approximations we have adopted. They are accurate and consistent
to leading order in the streaming and dynamic diffusion limits. They
are accurate to first order in $\beta$ in the static diffusion limit,
since we have had to retain all order $\beta$ terms in this limit
because they are of leading order in dynamic diffusion problems. Also note that if in a
given problem one never encounters the dynamic diffusion regime, it is
possible to drop more terms, as we discuss in \S~\ref{algorithm}.

The equations are easy to understand intuitively. The term
$-\lambda\nabla E$ in the momentum equation
(\ref{momentumconservation}) simply represents the
radiation force $\kr\vecF/c$, neglecting distinctions between the
comoving and laboratory frames which are smaller than leading order
in this equation. Similarly, the terms $\pm\kp(4\pi B - cE)$ and
$\pm\lambda(2\kp/\kr-1)\vecv\cdot\nabla E$ in the two energy equations
(\ref{gasenergy}) and (\ref{radenergy}) represent radiation absorbed minus
radiation emitted by the gas, and the work done by the
radiation field as it diffuses through the gas. The factor $(2\kp/\kr-1)$ arises because the term contains contributions both from the Newtonian work and from a relativistically-induced mismatch between emission and absorption. The term
proportional to $\kp E/c$ represents another relativistic correction
to the work, this one arising from boosting of the flux
between the lab and comoving frames. In the radiation energy equation
(\ref{radenergy}), the first term on the left hand side is the
divergence of the radiation flux, i.e.\ the rate at which radiation
diffuses, and the last term on the right hand side represents advection of the radiation enthalpy
$E+\calp$ by the gas.

It is also worth noting that equations (\ref{en2}) and (\ref{raden2})
are manifestly energy-conserving, since every term in one equation
either has an obvious counterpart in the other with opposite sign, or
is clearly an advection. In contrast, the momentum equation
(\ref{momentumconservation}) is not manifestly momentum-conserving, 
since there is a
force term $-\lambda\nabla E$ with no equal and opposite
counterpart. This non-conservation of momentum is an inevitable
side-effect of using the flux-limited diffusion approximation, since
this approximation amounts to allowing the radiation field to transfer
momentum to the gas without explicitly tracking the momentum of the
radiation field and the corresponding transfer from gas to radiation.

\section{The Importance of Higher Order Terms}
\label{termanalysis}

Our dynamical equations result from retaining at least some terms that
are formally of order $\beta^2$. Even though our analysis shows that
these terms can be the leading ones present, due to cancellations of
lower order terms, one might legitimately ask whether they are ever
physically significant. In \S~\ref{ordcomparison} we address this
question by comparing our equations to those that result from lower
order treatments. In \S~\ref{framecomparison}, we also compare our
equations with those generally used in comoving frame formulations of
radiation hydrodynamics.

To make our work in this section more
transparent, and since we are more interested in physical intuition
than rigorous derivation here, we specialize to the diffusion regime
in gray materials. Thus, we set $\lambda = R_2 = 1/3$ and
$\kp=\kr=\kz$. A more general analysis produces the same
conclusions, but is more mathematically cumbersome. We also focus on
the radiation energy equation, since all the terms that appear in the
gas energy equation also appear in it, and because there are no higher
order terms present in the momentum equation. Under these assumptions,
our radiation energy equation (\ref{radenergy}) becomes
\begin{eqnarray}
\frac{\partial}{\partial t}E
& = &
\nabla\cdot\left(\frac{c}{3 \kz} \nabla E\right) + \kz (4\pi B - c E)
\nonumber \\
& & {}
- \frac{4}{3}\nabla \cdot(\vecv E)
- \frac{1}{3} \vecv \cdot \nabla E 
+ \frac{4}{3}\kz\frac{v^2}{c} E.
\label{difffull}
\end{eqnarray}

\subsection{Comparison to Lower Order Equations}
\label{ordcomparison}

A common approach in radiation-hydrodynamic problems is to expand the
equations in $\beta$, rather than in both $\beta$ and $\tau$ as
we have done, and drop at least some terms that are of order $\beta^2$
in every regime \citep[e.g][]{mihalas99}. To determine how equations
derived in this manner compare to our higher order treatment, we
compare our simplified energy equation (\ref{difffull}) to the
corresponding equation one would obtain by following this procedure
with (\ref{radenergy}). This resulting energy equation is
\begin{eqnarray}
\frac{\partial}{\partial t} E & = & \nabla\cdot\left(\frac{c}{3\kz}
\nabla E \right) + \kz (4 \pi B - c E)
\nonumber \\
& &
{} - \frac{4}{3}\nabla\cdot(\vecv
E) - \frac{1}{3} \vecv\cdot\nabla E.
\label{difffirstorder}
\end{eqnarray}
\textit{It is important to caution at this point that, as we show
below, equation (\ref{difffirstorder}) is not accurate to leading
order in at least some cases, and should not be used for computations
unless one carefully checks that the missing terms never become
important in the regime covered by the computation.}

Compared to the energy equation (\ref{difffull}) that we obtain by
retaining all leading order terms in $\beta$ and $\tau$,
(\ref{difffirstorder}) is missing
the term $(4/3) \kz v^2 E/c$. If we think of the flux as having two
parts, a ``diffusion" part proportional to $\nabla E$ that comes from
radiation diffusion in the comoving frame, and a ``relativistic" part
proportional to $\vecv E + \vecv\cdot\calp$ that comes from the
Lorentz transformation between lab and comoving frames, then it is
natural to describe the $\vecv\cdot\nabla E$ term as the ``diffusion
work'' arising from the combination of the diffusion flux and the
post-Newtonian emission-absorption mismatch (as discussed in
\S~\ref{equationderivation}), and the $\kz v^2 E/c$ as the
``relativistic work'' arising from the relativistic flux. The presence
or absence of this relativistic work term is the difference between
our leading order-accurate equation and the equation one would derive
by dropping $\beta^2$ terms. Analyzing when, if ever, this term 
is physically important lets us identify in which situations a lower
order treatment may be inadequate.

If we use Table \ref{eqscalings} to compare the relativistic work term
to the
emission/absorption term, we find that $(\kz v^2 E/c) / [\kz (4\pi B -
c E)]$ is of order $\beta^2 \tau^2$ for static diffusion, and of order
unity
for dynamic diffusion. Thus, the term is never important in a static
diffusion problem, but is always important for a non-uniform,
non-equilibrium  dynamic diffusion problem system. We add the caveats
about non-uniformity and time-dependence because in a system where
there is no radiation-gas energy exchange, the relativistic work term
will be small due to a cancellation. The
example in Appendix \ref{dyndiffusionscaling} shows that in an
equilibrium, uniform medium, the terms $\kz (4\pi B-c E)$ and $(4/3)
\kz v^2 E/c$ cancel exactly at orders up to $\beta^2$. We expect any
system where variations occur on a scale for which $\beta\tau\gg 1$ to
resemble such a uniform, equilibrium medium, and thus we do not expect
the term $(4/3) \kz v^2 E/c$ to be important in such a system.

That said, there is still clearly a problem with omitting the
relativistic work term in a system where $\beta\tau \sim 1$. In this
case, Table \ref{eqscalings} implies that \textit{every} term on the
right hand side is roughly equally important regardless of whether we
use the static of dynamic diffusion scalings. To illustrate this
point, consider a radiation-dominated shock. The width of such a shock
is set by the balance
between radiation diffusing upstream from the hot post-shock region
into the cold pre-shock region, and advection of the radiation back
downstream by the pre-shock gas. This condition requires that
$\beta\tau\sim 1$ across the shock \citep{mihalas99}, so its width $w
\sim \lp/\beta$. Since $E$ changes by of order unity
across this distance, its spatial derivative is of order $\nabla E\sim
E/w\sim (\beta/\lp) E$. Applying this to (\ref{difffull}), we find
that each term on the right hand side is of order $\beta^2 (c/\lp)
E$. Since the terms like $-(4/3) \nabla \cdot (\vecv E)$ describing
advection and $\nabla\cdot[c/(3\kz)\nabla E]$ describing diffusion are
obviously important in the structure of the shock, causing order unity
changes in $E$, and the relativistic work term is comparable, it
follows that the relativistic work term is equally important. One can
obtain the correct structure within a radiation-dominated shock only
by retaining the relativistic work term.

An interesting point to note here is that omitting the relativistic
work term will not produce errors upstream or downstream of a shock,
because $\beta\tau \gg 1$ in these regions. Furthermore, the jump
conditions across a shock should be correct. The omitted term
affects radiation-gas energy exchange, not total energy conservation,
and all that is required to get the correct jump conditions are
conservation of mass and energy, plus correct computation of the
upstream and downstream radiation pressures. The lower order treatment
will therefore only make errors within the shock. Whether this is
physically important, or it is sufficient to get the jump conditions
correct, depends on whether one is concerned with structures on scales
for which $\beta\tau\sim 1$. An astrophysical example of a system
where one does care about structures on this scale is a
radiation-dominated accretion disk subject to photon bubble
instability \citep{turner03}. Such disks are in the dynamic diffusion
regime over the entire disk, but photon bubbles form on small scales
within them, and individual bubbles may have $\beta\tau\sim 1$ across
them.

\subsection{Comparison to Comoving Frame Formulations}
\label{framecomparison}

Many popular numerical treatments of radiation hydrodynamics
\citep[e.g.][]{turner01, whitehouse04, hayes06} use a comoving
formulation of the equations rather than our mixed frame
formulation. It is therefore useful to compare our equations to the
standard comoving frame equations. In the comoving formulation, the
evolution equation for the radiation field is usually the first law of
thermodynamics for the comoving radiation field \citep{mihalas82},
\begin{equation}
\label{firstlaw}
\rho \frac{D}{Dt}\left(\frac{E_0}{\rho}\right) +
\calp_0\colon(\nabla\vecv) = \kz (4 \pi B - cE_0) - \nabla \cdot \vecF_0.
\end{equation}
This equation is accurate to first order in $\beta$ in the sense that
it contains all the correct leading order terms and all terms that are
smaller than them by order $\beta$ or less.

To compare this to our mixed frame radiation energy equation
(\ref{radenergy}), we replace the comoving frame energy $E_0$ in
(\ref{firstlaw}) with the lab frame energy $E$ using the Lorentz
transformation (\ref{etransform}) and retain all terms that are of
leading order in any regime. In practice, this means that we set $E_0
= E$ inside the time derivative, since the difference between $E$ and
$E_0$ is at most $\beta/\tau$ or $\beta^2$ for static or dynamic
diffusion. However, when replacing $E_0$ with $E$ in the
heating/cooling term $4\pi B - c E_0$, we must retain all the terms in
(\ref{etransform}) because the leading term $4\pi B - c E$ is itself
only of order $\tau^{-2}$ or $\beta^2$ relative to $E$, so the
difference between $E$ and $E_0$ can be of leading
order.\footnote{Note 
that our need to retain difference between $E$ and $E_0$ here is
different from the situation when we first applied the Lorentz
transformation to derive (\ref{gztrans}) and
(\ref{gtrans}). In that case we did not need to retain the distinction
between $E$ and $E_0$, because in deriving
(\ref{gztrans}) and (\ref{gtrans}) there were no terms involving
$E_0$ explicitly. Instead, $E_0$ appeared only implicitly, as part of
the flux $\vecF$, and non-leading order corrections to $\vecF$ are not
of leading order in any regime. In contrast,
$E_0$ does appear explicitly in (\ref{firstlaw}).} This gives a
transformed equation
\begin{eqnarray}
\lefteqn{\rho \frac{D}{Dt}\left(\frac{E}{\rho}\right) +
\calp_0\colon(\nabla\vecv) =  \kz (4 \pi B - cE) - \nabla \cdot
\vecF_0
} \qquad\qquad
\nonumber \\
& & {} 
+ 2 \kz\frac{\vecv\cdot\vecF_0}{c} + \frac{\kz}{c} \left[ v^2 E +
(\vecv\vecv)\colon\calp_0\right].
\label{firstlaw1}
\end{eqnarray}

If we now adopt the diffusion approximation $\vecF_0 = -c/(3\kz)
\nabla E_0$ and $\calp_0=(1/3) E_0 \calI$, use the Lorentz
transformation to replace $E_0$ with $E$ throughout, and again only
retain terms that are of leading in order in some regime, then it is
easy to verify that (\ref{firstlaw1}) reduces to
(\ref{difffull}). Thus, our evolution equation is equivalent to the
comoving frame first law of thermodynamics for the radiation field,
\textit{provided that one retains all the leading order terms with
respect to $\beta$ and $\tau$, including some that are of order
$\beta^2$, when evaluating the Lorentz transformation}.

While the equations are equivalent, the mixed frame formulation has
two important advantages over the comoving frame formulation when it
comes to practical computation. First,
we are able to write the equations in a manner that allows a numerical
solution algorithm to conserve total energy to machine accuracy. We
present such an algorithm in \S~\ref{algorithm}. In contrast, it is
not possible to write a conservative update algorithm using the
comoving frame equations. The reason for this is that a conserved
total energy only exists in an inertial frame, and for a fluid whose
velocity is not a constant in space and time, the comoving frame is
not inertial. The lack of a conserved energy is a serious drawback to
comoving frame formulations.

A second advantage of the mixed-frame formulation is that it is far more
suited to implementation in codes with dynamically modified grid
structures such as adaptive mesh refinement methods. Since the
radiation energy is a conserved quantity, it is obvious how to refine
or coarsen it in a conservative manner. On the other hand, there is no
obviously correct method for refining or coarsening the comoving frame
energy density, because it will not even be defined in the same
reference frames before and after the refinement procedure.

\section{An Optimized Algorithm for Static Diffusion Radiation
Hydrodynamics}
\label{algorithm}

\subsection{Operator Splitting}

Our analysis shows that for static diffusion, the terms involving
diffusion and emission minus absorption of radiation always dominate
over those involving radiation work and advection. In addition, some
terms are always smaller than order $\beta$. This suggests an
opportunity for a significant algorithmic improvement over earlier
approaches while still retaining order $\beta$ accuracy in the
solution. In a simulation, one must update terms for the radiation
field implicitly, because otherwise stability requirements limit the
update time step to values comparable to the light-crossing time of a
cell. Standard approaches \citep[e.g.][]{turner01, whitehouse04,
whitehouse05, hayes06} therefore update all terms involving radiation
implicitly except the advection term and the radiation force term in
the gas momentum equation.

However, implicit updates are computationally expensive, so the
simpler the terms to be updated implicitly can be made, the simpler
the algorithm will be to code and the faster it will run. Since the
work and advection terms are non-dominant, we can produce a perfectly
stable algorithm without treating them implicitly. Even if this
treatment introduces numerically unstable modes in the work or
advection terms, they will not grow because the radiation diffusion
and emission/absorption terms, which are far larger, will smooth them
away each time step.

For the case of static diffusion, we therefore adopt the order $v/c$
equations (\ref{massconservation}) and (\ref{momentumconservation})
for mass and momentum conservation. For our energy equations, we adopt
(\ref{gasenergy}) and (\ref{radenergy}), but drop terms that are
smaller than order $\beta$ for static diffusion. This gives
\begin{eqnarray}
\frac{\partial}{\partial t}(\rho e) & = &
-\nabla[(\rho e + P)\vecv] - \kp (4\pi B - c E) 
\nonumber \\
& & {} + \lambda \left(2\frac{\kp}{\kr}-1\right)
\vecv\cdot\nabla E 
\label{gasenstatdiff}
\\
\frac{\partial}{\partial t}E & = &
\nabla\cdot\left(\frac{c\lambda}{\kr} \nabla E\right)
+ \kp (4\pi B - c E)
\nonumber \\
& & {}
-\lambda \left(2\frac{\kp}{\kr}-1\right) \vecv \cdot \nabla E
\nonumber \\
& & {}
- \nabla \cdot \left(\frac{3-R_2}{2}\vecv E\right)
\label{radenstatdiff}
\end{eqnarray}
To solve these, we operator split the diffusion and emission/absorption
terms, which we treat implicitly, from the work and advection terms,
which we treat explicitly. To do this, we write our gas/radiation
state as
\begin{equation}
\vecq = 
\left(
\begin{array}{c}
\rho \\
\rho \vecv \\
\rho e \\
E
\end{array}
\right),
\end{equation}
and our evolution equations as
\begin{equation}
\frac{\partial \vecq}{\partial t} 
= \vecf_{\rm e-nr} + \vecf_{\rm e-rad} + \vecf_{\rm i-rad},
\end{equation}
where we have broken our right hand side up into non-radiative terms
to be handled explicitly,
\begin{equation}
\vecf_{\rm e-nr} = 
\left(
\begin{array}{c}
-\nabla\cdot(\rho\vecv) \\
-\nabla\cdot(\rho\vecv\vecv)-\nabla P \\
-\nabla\cdot[(\rho e+P)\vecv] \\
0
\end{array}
\right),
\end{equation}
radiative terms to be handled explicitly,
\begin{equation}
\vecf_{\rm e-rad} = 
\left(
\begin{array}{c}
0 \\
-\lambda\nabla E \\
\lambda (2\frac{\kp}{\kr}-1) \vecv\cdot\nabla E \\
-\lambda(2\frac{\kp}{\kr}-1)\vecv\cdot\nabla E
- \nabla\cdot\left(\frac{3-R_2}{2}\vecv E\right)
\end{array}
\right),
\end{equation}
and radiative terms that must be handled implicitly,
\begin{equation}
\vecf_{\rm i-rad} = 
\left(
\begin{array}{c}
0 \\
\mathbf{0} \\
-\kp (4\pi B - cE) \\
\nabla\cdot\left(\frac{c\lambda}{\kr} \nabla E\right) 
+ \kp (4\pi B - cE)
\end{array}
\right).
\end{equation}

\subsection{Update Scheme}

For each update cycle, we start with the state $\vecq^n$ at the old
time. We first perform an implicit update to the
radiation and gas energy densities using $\vecf_{\rm i-rad}$. Any number of
methods are possible for this. For our implementation of this
algorithm in the Orion adaptive mesh refinement (AMR) code, we use
the method of \citet{howell03}, which we will not discuss in detail
here. To summarize, the algorithm involves writing the equations using
second order accurate spatial discretization and a time discretization
that limits to backwards Euler for large values of $\partial E/\partial
t$ (to guarantee stability) and to Crank-Nicolson when $\partial
E/\partial t$ is small (to achieve second order time accuracy).
This yields a matrix
equation for the radiation and gas energy densities at the new time,
which may be solved on both individual grids and over a hierarchy
of nested grids (as is necessary for AMR) using standard multigrid
techniques. The output of this procedure is an intermediate state
$\vecq^{n,*}$ which has been updated for $\vecf_{\rm i-rad}$.

Once the implicit update is done, we compute the ordinary hydrodynamic
update. As with the implicit update, this may be done using the
hydrodynamics method of one's choice. For our implementation, we use
the Godunov method described by \citet{truelove98}, \citet{klein99}, and
\citet{fisher02}. This update gives us $\vecq^{n,\dagger}$, the state
updated for $\vecf_{\rm i-rad}$ and $\vecf_{\rm e-nr}$. The only
modification we make to the standard update algorithm is to include a
radiation pressure term in the effective sound speed used to compute
the Courant condition. Thus, we take
\begin{equation}
c_{\rm eff} = \sqrt{\frac{\gamma P + (4/9) E (1 - e^{-\kr\Delta
x})}{\rho}}
\end{equation}
and set the time step to
\begin{equation}
\Delta t = C \frac{\Delta x}{\max(|\vecv|+c_{\rm eff})},
\end{equation}
where $\gamma$ is the ratio of specific heats for the gas, $C$ is the
Courant factor (usually 0.5), and the maximum is evaluated over all
cells. For AMR, this condition is applied independently on each level
$l$, and the time step is set using the values of $\Delta t^l$ in the
standard AMR manner \citep[e.g.][]{klein99}. The factor $(1 -
e^{-\kr\Delta x})$ gives us a means of interpolating between optically
thick cells, where radiation pressure contributes to the restoring
force and thus increases the effective signal speed, and optically
thin cells, where radiation does not provide any pressure.

Finally, we compute the force and advection terms
in $\vecf_{\rm e-rad}$. In our implementation we compute all of these
at cell centers using second order centered differences. For $\nabla
E$ this is
\begin{equation}
(\nabla E)^{n,*}_{i,j,k} =
\left(
\begin{array}{c}
\frac{E^{n,*}_{i+1,j,k}-E^{n,*}_{i-1,j,k}}{2\Delta x}, \\
\frac{E^{n,*}_{i,j+1,k}-E^{n,*}_{i,j-1,k}}{2\Delta y}, \\
\frac{E^{n,*}_{i,j,k+1}-E^{n,*}_{i,j,k-1}}{2\Delta z} \\
\end{array}
\right).
\end{equation}
Other derivatives are computed in an analogous manner. We then find
the new state by
\begin{equation}
\vecq^{n+1} = \vecq^{n,\dagger} + \vecf_{\rm e-rad} \Delta t.
\end{equation}

This update is manifestly only first order-accurate in time for the
explicit radiation terms, but there is no point in using a more
complex update because our operator splitting of some of the radiation
terms means that we are performing our explicit update using a
time-advanced radiation field, rather than the field at a half time
step. (\citealt{truelove98} show that one can avoid this problem for
gravitational body forces because the potential is linear in the
density, so it is possible to derive the half-time step potential from
the whole time step states. No such fortuitous coincidence occurs for
the radiation field.) This necessarily limits us to first order
accuracy in time for the terms we treat explicitly. However, since
these terms are always small compared to the dominant radiation terms,
the overall scheme should still be closer to second order than first
order in accuracy.

\subsection{Advantages and Limitations of the Method}

Our algorithm has two significant advantages in comparison to other
approaches, in particular those based on comoving frame formulations
of the equations \citep[e.g.][]{turner01, whitehouse05,hayes06}.
In any of these approaches,
since the radiation work terms are included in the
implicit update, one must solve an implicit quartic equation arising
from the combination of the terms $\kp(4\pi B - c E)$ and
$\calp\colon\nabla \vecv$. This may be done either at the same time
one is iterating to update the flux divergence term
$\nabla\cdot\vecF$ \citep{whitehouse05}, or in a separate iteration to
be done once the iterative solve for the flux divergence update is
complete \citep{turner01, hayes06}. In contrast, since our iterative
update involves only $\kp(4\pi B - c E)$ and $\nabla\cdot\vecF$, using
the \citet{howell03} algorithm we may linearize the equations and
never need to solve a quartic, leading to a simpler update algorithm
and a faster iteration step. Moreover, by using the \citet{howell03}
time-centering, we obtain second order accuracy in time whenever $E$
is changing slowly, as opposed to the backwards Euler differencing of
\citet{turner01}, \citet{whitehouse05}, and \citet{hayes06}, which is
always first order-accurate in time. Thus, our algorithm provides a
faster and simpler approach than the standard one.

A second advantage of our update scheme is that it retains the
total energy-conserving character of the underlying equations. In each
of the update steps involving radiation, for $\vecf_{\rm e-rad}$ and
$\vecf_{\rm i}$, the non-advective update terms in the radiation and
gas energy equations are equal and opposite. Thus, it is trivial to
write the update scheme so that it conserves total energy to machine
precision. This property is particularly important for turbulent
flows with large radiation energy gradients, such as those that occur
in massive star formation \citep[e.g.][]{krumholz07a}, because
numerical non-conservation is likely to be exacerbated by the
presence of these features. In contrast, in comoving frame formalisms
such as those of \citet{turner01}, \citet{whitehouse04}, and
\citet{hayes06} the exchange terms in their gas and
radiation energy equations are not symmetric. As a result, their
update schemes do not conserve total energy exactly. The underlying
physical reason for this asymmetry is that total energy is conserved
only in inertial frames such as the lab frame; it is not conserved in
the non-inertial comoving frame. For this reason, there is no easy way
to write a conservative update scheme from a comoving formulation.

Our algorithm also has two significant limitations, one obvious and
one subtle. The obvious limit is that our algorithm is only applicable
for static diffusion problems. For dynamic diffusion problems, e.g.\
stellar interiors or radiation-dominated shocks, our scheme is
unstable unless an appropriately small timestep is used. Whether this
instability is due to the explicit advection term, the explicit work
term, or both is not clear. Since codes such as ZEUS \citep{hayes06}
treat the advection explicitly without instability, however, it seems
likely that the work term is the culprit. Regardless of the
cause, even if we were to use a time step small enough to guarantee
stability, since the work and advection terms can be comparable to or
larger than the diffusion and heating/cooling terms for dynamic
diffusion, an algorithm that treats all the terms implicitly or all
explicitly, rather than our mix, is likely to be more accurate.

The subtle limitation is in our treatment of the hydrodynamics. We perform
the hydrodynamic update using a Riemann solver unmodified for the
presence of radiation force, work, and heating and cooling
terms. These terms should change the characteristic velocities of the
wave families in ways that depend on the radiation hydrodynamic regime
of the system. For example, in optically thick systems we should have
a radiation-acoustic mode rather than a simple sound wave, and in
optically thin systems where the radiation time scale is short
compared to the mechanical time scale, a gas may act as if it were
isothermal even if it has $\gamma\ne 1$. In some cases, failure to
modify the Riemann solver appropriately for these effects may produce
substantial errors, including a reduction in the order of accuracy of
the method from second to first \citep{pember93, lowrie01,
miniati06}. The severity of these effects for a given problem depends
the degree of stiffness of the radiation source terms. It should
also be noted that the other radiation diffusion methods most commonly
used for three-dimensional problems also suffer from this defect, so
this is not a comparative disadvantage of our method relative to
others.

\section{Tests of the Static Diffusion Algorithm}
\label{tests}

Here we describe five tests of our static diffusion
algorithm, done using our implementation of the algorithm in the Orion
AMR code, various
aspects of which are described in detail by \citet[multifluid
hydrodynamics]{puckett92},
\citet[hydrodynamics and gravity]{truelove98},
\citet[hydrodynamics and gravity]{klein99},
\citet[gravity]{fisher02}, 
\citet[radiation transport]{howell03}, 
\citet[sink particles]{krumholz04},
and \citet[magnetohydrodynamics]{crockett05}. For all of these tests
we use a single fluid with no magnetic fields and no self-gravity.

\subsection{Non-Equilibrium Marshak Wave}

As an initial check of the gas-radiation energy exchange in our code
in a case when radiation pressure is not significant and the gas is at
rest, we simulate the non-equilibrium Marshak wave problem. In this
problem, a zero-temperature, motionless, gaseous medium occupying all
space at $z>0$ is subject to a constant radiation flux $F_{\rm inc}
\hat{\mathbf{z}}$ incident on its surface at $z=0$. The gas is held
stationary, appropriate for early times before hydrodynamic motions
become significant. The medium is gray, with opacity $\kr=\kp=\kappa$,
and the constant-volume specific heat capacity of the gas is taken to
have the same $T^3$ dependence as that of the radiation, i.e. $c_v =
[\partial (e-v^2/2) / \partial T_g]_v = \alpha T_g^3$, where $T_g$ is
the gas temperature. The gas is not assumed to be in thermal
equilibrium with the radiation field, so the gas and radiation
temperatures may be different.

\citet{su96} give a semi-analytic solution to the time-dependent
behavior of the radiation energy density $E(z,t)$ and gas temperature
$T_g(z,t)$ for this problem. They introduce the dimensionless position
and time variables $x \equiv \sqrt{3}\kappa z$ and $\tau \equiv (4 a_R
c \kappa/\alpha) t,$ and the ``retardation'' parameter $\epsilon\equiv 4
a_R/\alpha$, and show that the dimensionless radiation energy density
\begin{eqnarray}
u(x,\tau) & \equiv & \left(\frac{c}{4}\right)
\left[\frac{E(z,t)}{F_{\rm inc}}\right] \\
& = & 1 - \frac{2\sqrt{3}}{\pi} 
\int_0^{1} d\eta\; e^{-\tau \eta^2}
\left\{\frac{\sin[x\gamma_1(\eta)+\theta_1(\eta)]}
{\eta\sqrt{3+4\gamma_1^2(\eta)}}\right\} 
\nonumber
\\ & & {}
- \frac{\sqrt{3}}{\pi} e^{-\tau} 
\int_0^1 d\eta\; \left(e^{-\tau/(\epsilon \eta)} \right.
\nonumber
\\ & & \quad {}
\left.\left\{\frac{\sin[x\gamma_2(\eta)+\theta_2(\eta)]}
{\eta(1+\epsilon\eta)\sqrt{3+4\gamma_2^2(\eta)}}\right\}\right),
\label{uintegral}
\end{eqnarray}
where
\begin{eqnarray}
\gamma_1(\eta) & = & \eta \sqrt{\epsilon + \frac{1}{1-\eta^2}}, \\
\gamma_2(\eta) & = &
\sqrt{(1-\eta)\left(\epsilon+\frac{1}{\eta}\right)},
\end{eqnarray}
and
\begin{equation}
\theta_n(\eta) =
\cos^{-1} \sqrt{\frac{3}{3+4\gamma_n^2(\eta)}}.
\end{equation}
The dimensionless gas energy density is
\begin{eqnarray}
v(x,\tau) & \equiv & \left(\frac{c}{4}\right) \left[\frac{a_R
T_g^4(z,t)}{F_{\rm inc}}\right] \\
& = & 
u(x,\tau) - \frac{2\sqrt{3}}{\pi}
\int_0^1 d\eta \; 
\left(
e^{-\tau(1-\eta^2)}
\right.
\nonumber
\\ & & \quad
\left.
\left\{
\frac{
\sin[x\gamma_3(\eta)+\theta_3(\eta)]
}{
\sqrt{
4-\eta^2+4\epsilon\eta^2(1-\eta^2)}
}
\right\}
\right)
\nonumber
\\ & & {}
+ \frac{\sqrt{3}}{\pi} e^{-\tau}
\int_0^1 d\eta \;
\left(
e^{-\tau/(\epsilon \eta)}
\right.
\nonumber
\\ & & \quad
\left.
\left\{
\frac{
\sin[x \gamma_2(\eta)+\theta_2(\eta)]
}{
\eta\sqrt{3+4\gamma_2^2(\eta)}
}
\right\}
\right),
\label{vintegral}
\end{eqnarray}
where
\begin{equation}
\gamma_3(\eta) =
\sqrt{(1-\eta^2)\left(\epsilon+\frac{1}{\eta^2}\right)}.
\end{equation}
Numerical evaluation of the integrals (\ref{uintegral}) and
(\ref{vintegral}) for $u$ and $v$ is not trivial
because the integrands perform an infinite number of oscillations
about zero as $\eta\rightarrow 0$. Correct computation of the result
when $\tau$ is small and $x$ is large requires careful numerical
analysis to ensure that the positive and negative contributions cancel
properly (J. Bolstad, 2007, in preparation).

We compare the properly computed semi-analytic results for $u$ and $v$
to a calculation performed with Orion using $\kappa=1$ cm$^{-1}$
and $\alpha=32 a_r/c$ (so $\epsilon=0.5$). The computational domain
goes from $0$ to $15$ cm (and thus to an optical depth $\kappa z=15$),
and is resolved by 100 equally-sized cells. For this test, since we
are comparing to a pure diffusion result, we set the flux limiter
$\lambda=1/3$ everywhere.

\begin{figure}
\plotone{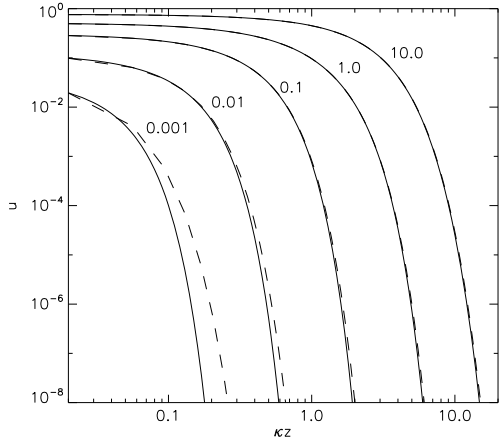} 
\caption{\label{marshakrad}
Dimensionless radiation energy density $u$ versus optical depth
$\kappa z$ at a series of times $\tau$. We show the semi-analytic
solution (\textit{solid lines}) and the solution computed with Orion
(\textit{dashed lines}). The value of the dimensionless time $\tau$ is
indicated by each curve.
}
\end{figure}

\begin{figure}
\plotone{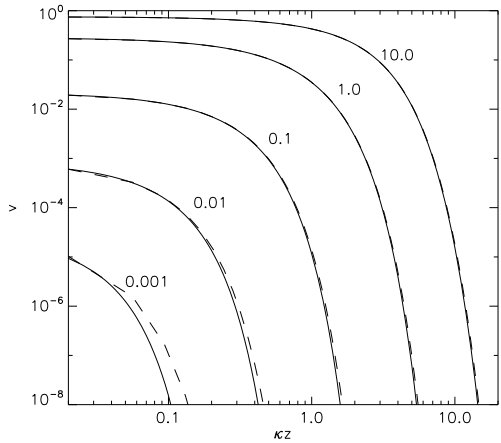} 
\caption{\label{marshakgas}
Same as Figure \ref{marshakrad}, but for the dimensionless gas energy
density $v$.
}
\end{figure}

Figures \ref{marshakrad} and \ref{marshakgas} compare the
semi-analytic dimensionless radiation and gas energy densities with
the values computed by Orion. At $\tau=0.001$ the agreement is fairly
poor due to low numerical resolution, since the wave only reaches an
optical depth of $\kappa z\sim 0.2$ and $\kappa z = 0.15$ is the size
of an individual computational cell. However, at later times when the
wave is resolved by a reasonable number of cells, the agreement
between the code result and the semi-analytic solution is excellent.

\subsection{Radiating Blast Wave}

We next compare to a test problem in which the gas is not at rest:
a Sedov-type blast wave with radiation
diffusion. \citet{reinicke91} gave the first similarity solution to
the problem of a point explosion with heat conduction, and following
\citet{shestakov99} and \citet{shestakov01}, we can adapt this
solution to the case of a point explosion with radiation
diffusion. This tests our code's ability to follow coupled
radiation-hydrodynamics in cases where radiation pressure is small.

We first summarize the semi-analytic solution. Consider an $n=3$
dimensional space filled with an adiabatic gas with equation of
state $P=(\gamma-1)\rho e \equiv \Gamma \rho T$, where $\Gamma$ is the
gas constant. The Planck mean opacity $\kp$ of the gas is very high,
so the gas and radiation temperatures are always equal. The Rosseland
mean opacity has a powerlaw form $\kr=\kappa_{0R,0} \rho^m T^{-n}$,
and we assume that it is always high enough to place us in the
diffusion regime, so $\lambda = 1/3$. Note that the choice of $-n=-3$
as the exponent of the opacity powerlaw is a necessary condition for
applying the \citet{reinicke91} conduction solution to our radiation
diffusion problem. Moreover, the similarity solution does not include
radiation energy density or pressure, so we consider only temperatures
for which the gas energy density and pressure greatly exceed the
radiation energy density and pressure, i.e. $\rho e \gg a_R
T^4$.

Under the assumptions described above, we may re-write the gas and
radiation energy equations (\ref{gasenergy}) and (\ref{radenergy}) as
a single conduction-type equation for the temperature,
\begin{equation}
\rho c_v \frac{\partial}{\partial t}T = \nabla(\chi_0 \rho^a T^b \nabla
T),
\end{equation}
where $c_v=\partial e/\partial T = \Gamma/(\gamma-1)$ is the
constant-volume specific heat of the gas, $\chi_0=4 c a_R/(3
\kappa_{0R,0})$, $a=-m$, and $b=n+3$. This equation has the same form
as the conduction equation considered by \citet{reinicke91}.

Consider now a point explosion at the origin of a spherically
symmetric region with an initial powerlaw density distribution
$\rho(r,t=0) = g_0 r^{-k_\rho}$. Initially the gas temperature $T$ and
pressure $P$ are negligible. The explosion occurs at the origin at time
zero, so the initial gas energy density is $(\rho e)(r,t=0)=E_0
\delta(\vecr)$. \citet{reinicke91} show that if the
initial density profile has a powerlaw index
\begin{equation}
\label{krhoeq}
k_{\rho} = \frac{(2b-1) n + 2}{2b - 2a + 1},
\end{equation}
then one may obtain a similarity solution via the change of variables
\begin{eqnarray}
\xi & = & \frac{r}{\zeta t^\alpha} \\
G(\xi) & = & \frac{\rho(r,t)}{g_0 r^{-k_{\rho}}} \\
U(\xi) & = & v(r,t)\frac{t}{\alpha r} \\
\Theta(\xi) & = & T(r,t) \Gamma \left(\frac{\alpha r}{t}\right)^2.
\end{eqnarray}
Here, $\xi$, $G(\xi)$, $U(\xi)$, and $\Theta(\xi)$ are the
dimensionless distance, density, velocity, and temperature,
\begin{equation}
\alpha = \frac{2b - 2a + 1}{2b - (n+2)a + n},
\end{equation}
and $\zeta$ is a constant with units of
$[\mbox{length}][\mbox{time}]^{-\alpha}$ 
whose value is determined by a procedure we discuss below.

With this similarity transformation, the equations of motion and heat
conduction reduce to
\begin{eqnarray}
U' - (1-U)(\ln G)' + (n-k_{\rho}) U & = & 0 \\
(1-U) U' + U(\alpha^{-1}-U) & = &
\nonumber \\
\Theta[\ln (\xi^{2-k_{\rho}} G\Theta)]',
\end{eqnarray}
and
\begin{eqnarray}
\lefteqn{
2[U' + nU - \mu (\alpha^{-1}-1)]
=
\mu (1-U) [\ln(\xi^2\Theta)]'
}
\nonumber \\
& &
{}
+ \beta_0 \Theta^b G^{a-1} \xi^{(2b-1)/\alpha} 
\cdot
\left((\ln \Theta)'' +
[\ln (\xi^2\Theta )]'
\right.
\\
& &
\left.
{}\cdot
 \left\{n-2+a[\ln(\xi^{-k_{\rho}} G)]' 
+
(b+1)
[\ln (\xi^2 \Theta)]'\right\}\right),
\end{eqnarray}
where $()'\equiv d()/d\ln \xi$, $\mu=2/(\gamma-1)$, and
\begin{equation}
\beta_0 = \frac{2\chi_0 (\alpha \zeta^{1/\alpha})^{2b-1}}{\Gamma^{b+1}
g_0^{1-\alpha}} \mbox{sgn}(t).
\end{equation}
This constitutes a fourth-order system of non-linear ordinary
differential equations. All physical solutions to these equations pass
through two discontinuities, a heat front and a shock front, with the
heat front at larger radius. However, the jump conditions for these
discontinuities are easy to determine, and one can integrate between
them. For a given $\beta_0$, the solution depends
only on the dimensionless parameter
\begin{equation}
\Omega = \frac{2\chi_0}{\Gamma^{b+1} g_0^{1-a}}
\left(\frac{E_0}{g_0}\right)^{b-1/2},
\end{equation}
which measures the strength of the explosion. Large values of $\Omega$
constitute ``strong'' explosions, and the ratio of heat front radius
to shock front radius is a monotonically increasing function of
$\Omega$. It is important at this point to add
a cautionary note: in deriving the similarity solution, we assumed
that radiation energy density is negligible in comparison to gas
energy density. This cannot strictly be true at early times, since at
$t=0$ the temperature diverges at the origin, and the radiation energy
density varies as $T$ to a higher power than the gas energy
density. However, the true behavior should approach the similarity
solution at later times.

While we have reduced the gas dynamical equations to a system of ordinary
differential equations that is trivial to integrate, solving the full
problem is complex because the equations still depend on the unknown
parameter $\beta_0$, which in turn depends on
$\zeta$. To solve the problem, we must determine $\beta_0$ from the
given initial conditions.
\citet{reinicke91} describe the iteration procedure required
to do this in detail, and we only summarize it here. To find a
solution, one first chooses a value $\xi_h>1$ for the dimensionless
radius of the heat front, applies the boundary conditions at the front, and
guesses a corresponding value of $\beta_0$. For each $\xi_h$ there
exists a unique $\beta_0$ for which it is possible
to integrate the equations back from $\xi=\xi_h$ to the location of
the shock front at $\xi=\xi_s$, apply
the shock jump conditions, and continue integrating back to the origin
at $\xi=0$ without having the solution become double-valued and thus
unphysical. One iterates to identify the allowed value of
$\beta_0$ for the chosen $\xi_h$, and this gives the unique density,
velocity, and temperature profiles allowed for that $\xi_h$.
However, the solution one finds in
this way may not correspond to the desired value of $\Omega$.
\citeauthor{reinicke91} show that
\begin{equation}
\Omega = \beta_0 \left[2\pi \int_0^{\xi_h} \xi^{n-k_{\rho}+1} G (U^2 +
\mu \Theta)\, d\xi\right]^{b-1/2}.
\end{equation}
Thus, each choice of $\xi_h$ corresponds to a particular value of
$\Omega$, and one must iterate a second time to find the value of
$\xi_h$ that gives the value of $\Omega$ determined from the input
physical parameters of the problem. Alternately, instead of specifying
a desired value of $\Omega$, one may specify a ratio $R=\xi_h/\xi_s$,
which also determines a unique value for $\xi_h$.

For our comparison between the semi-analytic solution and Orion, we
adopt the parameters $\gamma=7/5$,
$c_v=1/(\gamma-1)$, $a=-2$, $b=6$, $g_0=\chi_0=1$, and $E_0 = 135$,
which yields a strength $\Omega = 1.042\times 10^{12}$ and a ratio
$R=2.16$. In the simulation, we turn off terms
in the code involving radiation pressure and forces, and we set
$\lambda=1/3$ exactly. We use one-dimensional spherical polar
coordinates rather than Cartesian coordinates; the solution procedures
for this are identical to the ones outlined in \S~\ref{algorithm},
with the exception that the gradient and divergence operators have
their spherical rather than Cartesian forms, and the cell-centered
finite differences are modified appropriately. Our computational
domain goes covers $0 \leq r \leq 1.05$, resolved by 256, 512, or 1024
cells, and has reflecting inner and outer boundary conditions. To
initialize the problem we set initial density to the powerlaw profile
$\rho=r^{-k_{\rho}}$ (with $k_{\rho}$ set from equation \ref{krhoeq}),
the initial velocity to zero, and the initial energy density to a
small value, except in the cell adjacent to the origin, where its
value is $\rho e = 135/(\gamma-1)$.

\begin{figure}
\plotone{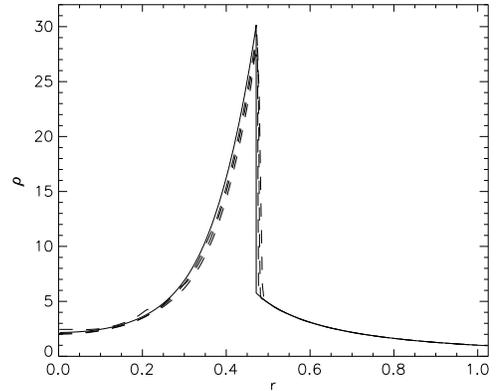}
\caption{\label{blastwaveden}
Density $\rho$ versus radius $r$ for the
radiating blast wave test. We show the semi-analytic solution
(\textit{solid line}), and the Orion results at resolutions of 256,
512, and 1024 cells (\textit{dashed lines}). The 256-cell run is the
dashed line furthest from the semi-analytic solution, and the
1024-cell run is the dashed line closest to it.
}
\end{figure}

\begin{figure}
\plotone{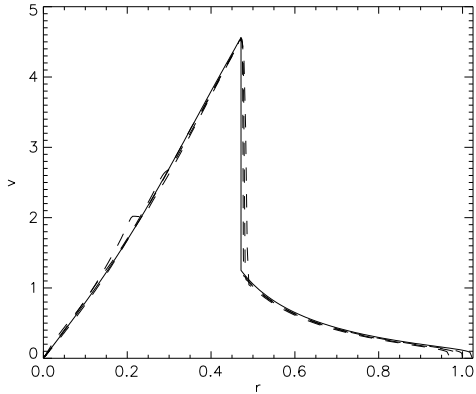}
\caption{\label{blastwavevel}
Same as Figure \ref{blastwaveden}, but for the velocity
$v$.
}
\end{figure}

\begin{figure}
\plotone{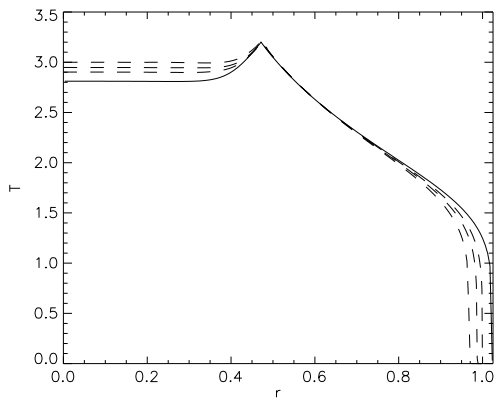}
\caption{\label{blastwavetemp}
Same as Figure \ref{blastwaveden}, but for the temperature
$T$.
}
\end{figure}

Figures \ref{blastwaveden}, \ref{blastwavevel}, and
\ref{blastwavetemp} compare the semi-analytic density, velocity, and
temperature profiles to the values we obtain from Orion after
running to a time $t=0.06$. As the plots show, the Orion results agree
very well with the semi-analytic solution, and the agreement improves
with increasing resolution. In the lowest resolution run, there is a
small oscillation in the density and velocity about a third of the way
to the shock, which is likely due to the initial blast energy being
deposited in a finite-volume region rather than as a true $\delta$
function. However, this vanishes at higher resolutions. Overall, the
largest errors are in the temperature in the shocked gas.

\begin{figure}
\plotone{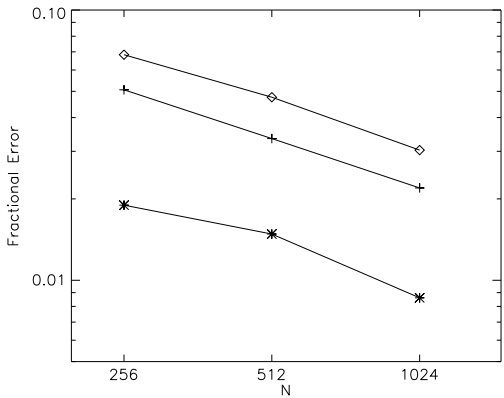}
\caption{\label{blastwaveerr}
Fractional error versus resolution $N$ in the radiating blast wave
test. The fractional error is defined as $|\mbox{simulation value} -
\mbox{analytic value}|/\mbox{analytic value}$. We show error in the
heat front radius $r_h$ (\textit{plus signs}), shock front radius
$r_s$ (\textit{asterisks}), and their ratio $R=r_h/r_s$
(\textit{diamonds}).
}
\end{figure}

As a metric of convergence, we plot the error of our simulation
relative to the analytic solution as a function of resolution in
Figure \ref{blastwaveerr}. We do this for the quantities $r_h$ and
$r_s$, the positions of the shock and heat fronts, and their ratio
$R$. For this purpose, we define the location of the heat and shock
fronts for the simulations as the positions of the cell edges where
$dT/dr$ and $d\rho/dr$ are most negative. As the plot shows, at the
highest resolution the errors in all three quantities are $\ltsim
3\%$, and the calculation appears to be converging. The order of
convergence is roughly 0.6 in all three quantities. It is worth noting
that computing the locations of the heat and shock fronts is a
particularly strong code test, because obtaining the correct
propagation velocities for the two fronts requires that the code
conserve total energy very well. Non-conservative codes have
significant difficulties with this test \citep{timmes06}.

\subsection{Radiation Pressure Tube}
\label{radtube}

Our third test is to simulate a tube filled with radiation and 
gas. The gas within the tube is optically thick, so the diffusion
approximation applies. The two ends of the tube are held at fixed
radiation and gas temperature, and radiation diffuses through the
gas from one end of the tube to the other. The radiation flowing
through the tube exerts a force on the gas, and the gas density
profile is such that, with radiation pressure, the gas is in pressure
balance and should be stationary. For computational simplicity, we set
the Rosseland- and Planck-mean opacities per unit mass of the gas to
a constant value $\kappa$. A simulation of this system tests our
code's ability to compute accurately the radiation pressure force in
the very optically thick limit.

We first derive a semi-analytic solution for the configuration
of the tube satisfying our desired conditions. Since the gas is very
optically thick and we are starting the system in equilibrium, we set
$T_{\rm rad} = T_{\rm gas} \equiv T$. The fluid is initially at
rest. The condition of pressure balance amounts to setting $\partial (\rho
\vecv)/\partial t + \nabla\cdot(\rho\vecv\vecv)=0$ 
in equation (\ref{momentumconservation}), so that the
radiation pressure force balances the gas pressure gradient. Thus, we have
\begin{eqnarray}
\frac{dP}{dx} + \lambda \frac{dE}{dx} & = & 0 \\
\left(\frac{k_B}{\mu} \rho + \frac{4}{3} a_R T^3\right) \frac{dT}{dx} +
\frac{k_B}{\mu} T \frac{d\rho}{dx} & = & 0.
\label{radtube1}
\end{eqnarray}
In the second step we have set $E=a_R T^4$ and $P=\rho k_B T/\mu$, where
$\mu$ is the mean particle mass, and we have set $\lambda=1/3$ as is
appropriate for the optically thick limit. The radiation energy
equation (\ref{radenstatdiff}) for our configuration is simply
\begin{eqnarray}
\frac{d}{dx} \left( \frac{c \lambda}{\kappa \rho} \frac{dE}{dx}\right)
& = & 0 \\
\frac{d^2 T}{dx^2} + 3\frac{1}{T} \left(\frac{dT}{dx}\right)^2 -
\frac{1}{\rho} \left(\frac{d\rho}{dx}\right)
\left(\frac{dT}{dx}\right) & = & 0.
\label{radtube2}
\end{eqnarray}

\begin{figure}
\plotone{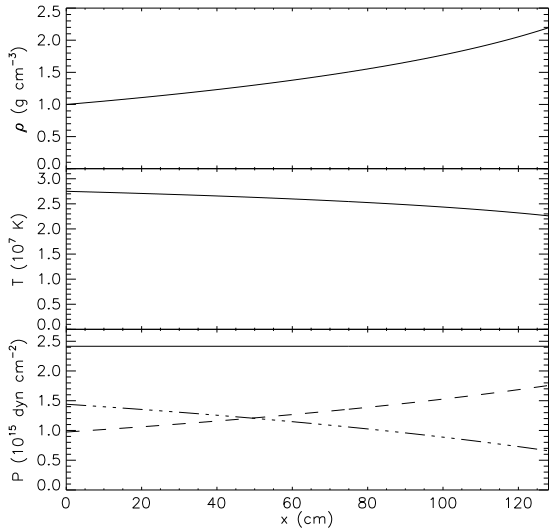}
\caption{\label{radtubesol}
Density, temperature, and pressure versus position in the radiation
tube problem. The bottom panel shows total pressure (\textit{solid
line}), gas pressure (\textit{dashed line}), and radiation pressure
(\textit{dot-dashed line}).
}
\end{figure}

Equations (\ref{radtube1}) and (\ref{radtube2}) are a pair of coupled
non-linear ordinary differential equations for $T$ and $\rho$. The
combined degree of the system is three, so we need three initial
conditions to solve them. Thus, let the tube run from $x=x_0$ to
$x=x_1$, with temperature, density, and density gradient $T_0$,
$\rho_0$, and $(d\rho/dx)_0$ at $x_0$. For a given choice of initial
conditions, it is trivial to solve (\ref{radtube1}) and
(\ref{radtube2}) numerically to find the density and temperature
profile. We wish to investigate both the radiation pressure and gas
pressure dominated regimes, so we choose parameters to ensure
that our problem covers both. The choice $x_0=0$, $x_1=128$ cm,
$\rho_0=1$ g cm$^{-3}$, 
$(d\rho/dx)_0 = 5\times 10^{-3}$ g cm$^{-4}$, and $T_0=2.75 \times 10^7$ 
K satisfies this requirement if we adopt $\mu=2.33 \,m_{\rm P}=3.9\times
10^{-24}$ g and $\kappa=100$ cm$^2$ g$^{-1}$. Figure \ref{radtubesol} shows the
density, temperature, and pressure as a function of position for these 
parameters.

\begin{figure}
\plotone{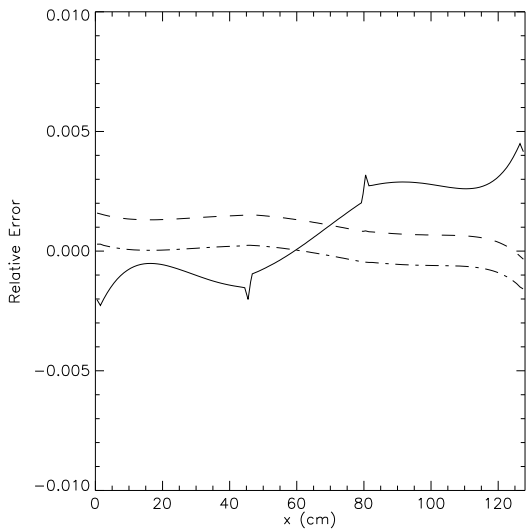}
\caption{\label{radtubeerr}
Relative error in density (\textit{solid line}), gas temperature
(\textit{dashed line}), and radiation temperature (\textit{dot-dashed
line}) in the radiation tube test.
}
\end{figure}

We solve the equations to obtain the density and temperature as a
function of position, and then set these values as initial conditions
in a simulation. The simulation has 128 cells along the length of the
tube on the coarsest level. We impose Dirichlet boundary conditions on the
radiation field, with the radiation temperature at each end of the tube set
equal to its value as determined from the analytic solution. We use
symmetry boundary conditions on the hydrodynamics, so that gas can
neither enter nor leave the computational domain. To ensure that our
algorithm does not 
encounter problems at the boundaries between AMR levels, we refine
the central $1/4$ of the problem domain to double the resolution of
the base grid. We evolve the system for 10 sound crossing times and
measure the amount by which the density and temperature change
relative to the exact solution. We plot the relative error, defined as 
(numerical solution $-$ analytic solution) / (analytic solution), in the
density, gas temperature, and radiation temperature in Figure
\ref{radtubeerr}. As the plot shows, our numerical solution agrees
with the analytic result to better than $0.5\%$ throughout the
computational domain. The density error is smallest in the higher 
resolution central region, as expected. There is a very small increase 
in error at level boundaries, but it is still at the less than $0.5\%$
level.

\subsection{Radiation-Inhibited Bondi Accretion}

The previous test focuses on radiation pressure forces in the
optically thick limit. To test the optically thin limit, we
simulate accretion onto a radiating point particle. We consider a
point mass $M$ radiating with a constant luminosity $L$ accreting
from a background medium. The medium consists of gas which has zero
velocity and density $\rho_{\infty}$ far from the particle. We take
the gas to be isothermal with constant temperature $T$, and enforce
that it is not heated or cooled radiatively by setting its Planck
opacity $\kp=0$. We set the Rosseland opacity of the gas to a constant
non-zero value $\kr$, and choose $\rho_{\infty}$ such that the
computational domain is optically thin. In this case, the radiation
free-streams away from the point mass, and the radiation energy
density and radiative force per unit mass on the gas are
\begin{eqnarray}
\label{eanalyt}
E &= & \frac{L}{4\pi r^2 c} \\
\vecf_{\rm r} & = & \frac{\kr L}{4\pi r^2 c}
\left(\frac{\vecr}{r}\right),
\end{eqnarray}
where $\vecr$ is the radial vector from the particle and $r$ is its
magnitude. The gravitational force per unit mass is $\vecf_{\rm g}=- (G
M/r^2) (\vecr/r)$, so the net force per unit mass is
\begin{equation}
\vecf = \vecf_{\rm r} + \vecf_{\rm g} = -(1 - f_{\rm Edd}) \frac{G M}{r^2}
\left(\frac{\vecr}{r}\right),
\end{equation}
where
\begin{equation}
f_{\rm Edd} = \frac{\kr L}{4\pi G M c}
\end{equation}
is the fraction of the Eddington luminosity with which the point mass
is radiating.

Since the addition of radiation does not alter the $1/r^2$ dependence
of the specific force, the solution is simply the standard
\citet{bondi52} solution, but for an effective mass of $(1-f_{\rm
Edd}) M$. The accretion rate is the Bondi rate
\begin{equation}
\dot{M}_B = 4\pi \xi r_B^2 c_s \rho_{\infty},
\end{equation}
where
\begin{equation}
r_B = (1-f_{\rm Edd}) \frac{G M}{c_s^2}
\end{equation}
is the Bondi radius for the effective mass, $c_s$ is the gas sound
speed at infinity, and $\xi$ is a numerical factor of order unity that
depends on the gas equation of state. For an isothermal gas,
$\xi=e^{3/2}/4$, and the radial profiles of the non-dimensional
density $\alpha\equiv \rho/\rho_{\infty}$ and velocity $u\equiv v/c_s$
are given by the solutions to the non-linear algebraic equations
\citep{shu92}
\begin{eqnarray}
x^2 \alpha u & = & \xi \\
\frac{u^2}{2} + \ln \alpha - \frac{1}{x} & = & 0,
\end{eqnarray}
where $x\equiv r/r_B$ is the dimensionless radius.

To set up this test, we make use of the Lagrangian sink particle
algorithm of \citet{krumholz04}, coupled with the ``star particle''
algorithm of \citet{krumholz07a} which allows the sink particle to act
as a source of radiation. We refer readers to those papers for details
on the sink and star particle algorithms. We simulate a computational
domain $5\times 10^{13}$ cm on a side, resolved by $256^3$ cells, with a
particle of mass $M=10$ $\msun$ and luminosity $L=1.6\times 10^{5}$
$\lsun$ at its center. We adopt fluid properties $\rho_{\infty} =
10^{-18}$ g cm$^{-3}$, $\kr=0.4$ cm$^2$ g$^{-1}$, and $c_s=1.3\times
10^{7}$ cm s$^{-1}$, corresponding to a gas of pure, ionized hydrogen
with a temperature of $10^6$ K. With these values, $f_{\rm Edd} = 0.5$,
$r_B = 4.0\times 10^{12}$ cm, and $\dot{M}_B = 2.9 \times 10^{17}$ g
s$^{-1}$. We use inflow boundary conditions on the gas and Dirichlet
boundary conditions on the radiation field, with the radiation energy
density on the boundary set to the value given by equation
(\ref{eanalyt}).

\begin{figure}
\plotone{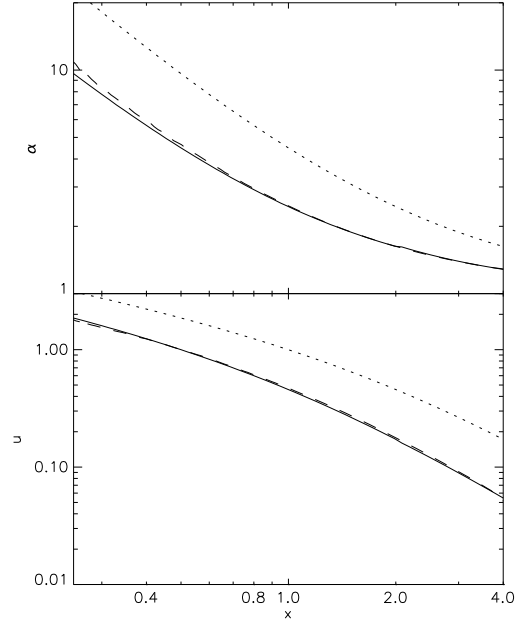}
\caption{\label{bondisol}
Dimensionless density $\alpha$ (\textit{upper panel}) and velocity $u$
(\textit{lower panel}) versus dimensionless position $x$ for
radiation-inhibited Bondi accretion. We show the analytic solution
(\textit{solid line}), the solution as computed with Orion
(\textit{dashed line}), and the analytic solution for Bondi accretion
without radiation (\textit{dotted line}). For the Orion result, the
values shown are the radial averages computed in $128$
logarithmically-spaced bins running from the accretion radius $x=0.25$
to the outer edge of the computational domain $x=5$.
}
\end{figure}

Figure \ref{bondisol} compares the steady-state density $\alpha$ and
velocity $u$ computed by Orion to the analytic solution. The agreement
is excellent, with differences between the analytic and numerical
solutions of $\sim 1\%$ everywhere except very near the accretion
radius at $x=0.25$. The maximum error is $\sim 10\%$ at the surface of
the accretion region; this is comparable to the error in density for
non-radiative Bondi accretion with similar resolution in
\citet{krumholz04}. In comparison, the solution is nowhere near the
solution that would be obtained without radiation. After running for
$5$ Bondi times ($=r_B/c_s$), the average accretion rate is $2.4
\times 10^{17}$ g s$^{-1}$. While this differs
from the analytic solution by $19\%$, the error is also not
tremendously different from that obtained by \citet{krumholz04} when
the Bondi radius was resolved by 4 accretion radii, and is nowhere
near the value of $1.2\times 10^{18}$ g s$^{-1}$ which would occur
without radiation.

We should at this point mention one limitation of our algorithm, as
applied on an adaptive grid, that this test reveals. The $1/r^2$
gradient in the radiation energy density is very steep, and we compute
the radiation force by computing gradients in $E$. We found that, in
an AMR calculation, differencing this steep gradient across level
boundaries introduced significant artifacts in the radiation pressure
force. With such a steep gradient, we were only able to compute the
radiation pressure force accurately on fixed grids, not adaptive
grids. This is not a significant limitation for most applications
though, since for any appreciable optical depth the gradient will be
much shallower than $1/r^2$. As the radiation pressure tube test in
\S~\ref{radtube} demonstrates, in an optically thick problem the
errors that arise from differencing across level boundaries are less
than 1\%.

\subsection{Advecting Radiation Pulse}

The previous two tests check our ability to compute the radiation
pressure force accurately in the optically thick
and optically thin limits. However, they do not strongly test
radiation advection by gas. To check this, we simulate a diffusing,
advecting radiation
pulse. The initial condition is a uniform background of gas and
radiation far from the pulse. Centered on $x=0$ there is an increase
in the radiation energy density and a corresponding decrease in the
gas density, so that the initial condition is everywhere in pressure
balance. As radiation diffuses out of the pulse, pressure support is
lost and the gas moves into the lower density region. We cannot solve
this problem analytically, but we can still perform a very useful
test of the methodology by comparing a case in which the gas is
initially at rest with respect to the computational grid with a case
in which the gas is moving at a constant velocity with respect to the
grid. The results should be identical when shifted to lie on top of
one another, but the work and advection terms will be different in the
stationary case than in the advected case. Checking that the results
do not change when we advect the problem enables us to determine if
our code is correctly handling the advection of radiation by the gas.

\begin{figure}
\plotone{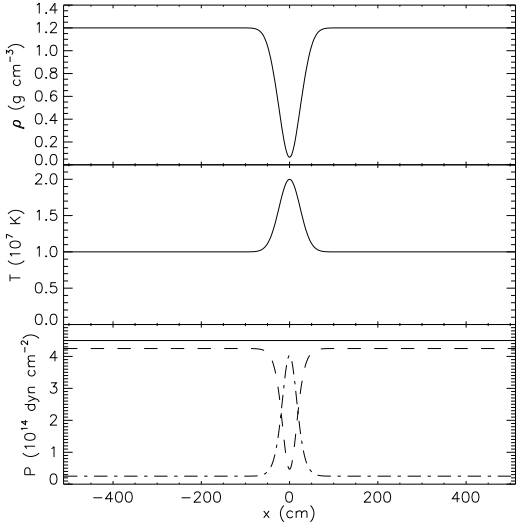}
\caption{\label{radadvectsetup}
Density, temperature, and pressure versus position in the advected
radiation pulse problem. The bottom panel shows total pressure (solid
line), gas pressure (dashed line), and radiation pressure (dot-dashed
line).
}
\end{figure}

For our simulations, we use equal initial gas and radiation
temperatures, with temperature and density profiles
\begin{eqnarray}
T & = & T_0 + (T_1-T_0) \exp\left(-\frac{x^2}{2 w^2}\right) \\
\rho & = & \rho_0 \frac{T_0}{T}+ \frac{a_R \mu}{3 k_B} \left(\frac{T_0^4}{T} -
T^3\right),
\end{eqnarray}
with $T_0 = 10^7$ K, $T_1=2\times 10^7$ K, $\rho_0 = 1.2$ g cm$^{-3}$,
$w = 24$ cm, $\mu=2.33 \,m_{\rm P}=3.9\times 10^{-24}$ g, and
$\kappa=100$ cm$^2$ g$^{-1}$. The
density, temperature, and pressure profiles are shown in Figure
\ref{radadvectsetup}. In the bottom panel, the solid line is the
total pressure, the dashed line is the gas pressure, and the
dot-dashed line is the radiation pressure. 
As the figure indicates, the system is initially 
in pressure balance. 

\begin{figure}
\plotone{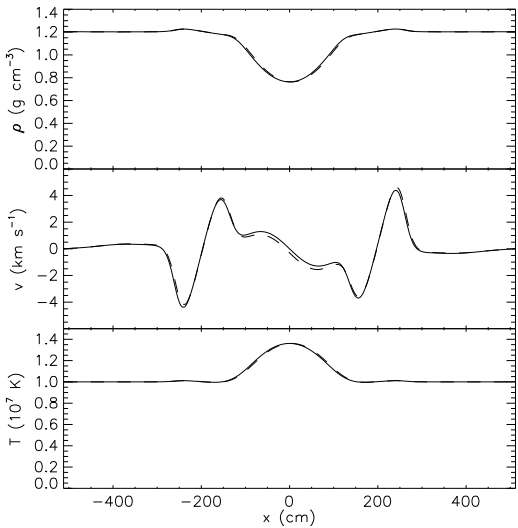}
\caption{\label{radadvectfinal}
Density, velocity, and temperature in the advected
radiation pulse problem after $4.8\times 10^{-5}$ s of evolution. In each panel the solid line is the unadvected run, and the dashed line is the advected run shifted 48 cm in the $-x$ direction. In the velocity plot, the velocity we show for the advected run is relative to the overall systematic velocity of $10^6$ cm s$^{-1}$ in the initial condition.
}
\end{figure}

\begin{figure}
\plotone{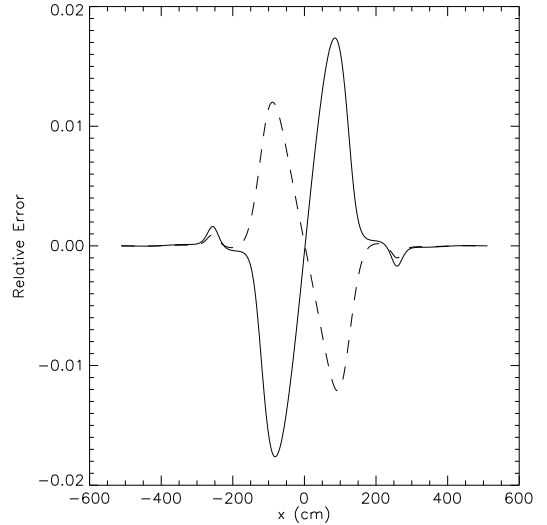}
\caption{\label{radadvecterr}
Relative error in density (solid line) and gas/radiation temperature
(dashed line) in the radiation pulse test.
}
\end{figure}

We compare two runs, one where the velocity
is zero everywhere and another with a uniform initial velocity
$v=10^6$ cm s$^{-1}$ in the $x$ direction. In both runs the simulation domain
extends from $-512$ to $512$ cm, resolved by 512 cells with no
adaptivity. We use periodic boundary
conditions on the gas and the radiation, and run for $4.8\times
10^{-5}$ s, long enough for the pulse to have been advected over its
own initial width twice.

To check our results, we shift the advected run by 48 cm in the $-x$
direction, so that it should lie on top of the unadvected run. Figure
\ref{radadvectfinal} shows the configuration of the advected and
unadvected runs at this point.
We then plot the relative difference between the advected
and unadvected runs, defined as $(\mbox{unadvected} - \mbox{advected})
/ \mbox{unadvected}$,
in Figure \ref{radadvecterr}. We do not differentiate between the gas and
radiation temperatures, because they are identical at the $10^{-3}$
level. We do not plot the error in velocity because the velocities in
the unadvected run are close to zero over most of the computational
domain. As the plot shows, the difference between the advected and
unadvected runs is less than $2\%$ everywhere in the simulation.

\section{Summary}
\label{summary}

We derive the correct equations for mixed frame flux-limited
diffusion radiation hydrodynamics. The error in our equations if of
order $v^2/c^2$ in the static
diffusion limit, and of order $v/c$ in the dynamic diffusion and
streaming limits. We give the equations in a form that is well-suited
to implementation in numerical simulations, because they make it
trivial to maintain exact conservation of total energy. Our analysis
reveals that lower order formulations of the equations, which neglect
differences between the laboratory and comoving frames, are incorrect at
order unity for systems in the dynamic diffusion limit. It remains to
be seen how serious this defect is in practice, but analytic arguments
suggest that at a minimum one ought to be very careful in applying
zeroth order codes to problems where there are interesting or
important structures on scales for which $\beta\tau \sim 1$. We give
the equations that are correct to leading order for dynamic diffusion,
which do not suffer from this problem.

Our analysis also reveals that, for static diffusion problems, one can
obtain a significant algorithmic simplification and speedup compared to
algorithms based on comoving frame formulations of the equations by
treating non-dominant radiation terms explicitly rather than
implicitly. This advance is possible even though the underlying
equations of our method conserve total energy to machine precision while
comoving frame formulations of the equations do not. This property is
particularly important for flows that are turbulent or otherwise
involve large gradients in gas or radiation properties, since these
are the problems most likely to suffer from numerical non-conservation. We
demonstrate an implementation of this method in the Orion adaptive
mesh refinement code, and show that it provides excellent agreement
with analytic and semi-analytic solutions in a series of test
problems covering a wide range of radiation-hydrodynamic regimes.

\acknowledgements We thank J.~I. Castor and
J.~M. Stone for helpful comments on the manuscript, and T.~A. Thompson
for helpful discussions. We also thank the referee for comments that
improved the paper. Support for this work was
provided by NASA through Hubble Fellowship grant \#HSF-HF-01186
awarded by the Space Telescope Science Institute, which is operated by
the Association of Universities for Research in Astronomy, Inc., for
NASA, under contract NAS 5-26555 (MRK); NASA ATP grants NAG 5-12042
and NNG06GH96G
(RIK and CFM); the US Department of Energy at the Lawrence Livermore
National Laboratory under contract W-7405-Eng-48 (RIK and JB); and the NSF
through grants AST-0098365 and AST-0606831 (CFM). This research was
also supported by grants of high performance computing resources from
the Arctic Region Supercomputing Center; the NSF San Diego
Supercomputer Center through NPACI program grant UCB267; the National
Energy Research Scientific Computing Center, which is supported by the
Office of Science of the U.S. Department of Energy under Contract
No. DE-AC03-76SF00098, through ERCAP grant 80325; and the US
Department of Energy at the Lawrence Livermore National Laboratory
under contract W-7405-Eng-48.


\begin{appendix}

\section{Scalings in the Dynamic Diffusion Limit}
\label{dyndiffusionscaling}

Here we show that the emission minus absorption term $4\pi B/c - E$
is of order $\beta^2 E$ in the dynamic diffusion
limit. \citet{mihalas99} argue that in this limit $4\pi B/c - E$ is of
order $(\beta/\tau)E$. However, this conclusion is based on their
analysis of the second order equilibrium diffusion approximation
\citep[pg. 461-466]{mihalas99}, in which they retain terms of order
$\beta/\tau$ while dropping those of order $\beta^2$. While this is
correct for static diffusion, in the dynamic diffusion limit $\beta^2
\gg \beta/\tau$, so the approach in \citeauthor{mihalas99} is not
consistent, and is insensitive to terms of order $\beta^2$.

We will not give a general proof that $4\pi B/c - E\sim \beta^2 E$ for
dynamic diffusion, but we can establish it by a simple thought
experiment. Consider a system that is infinitely far into the dynamic
diffusion limit, in the sense that $\tau=\infty$: an infinite
uniform medium that is at rest and in perfect thermal equilibrium
between the radiation field and the gas. In the rest frame of the
medium, these assumption require $E_0=4\pi B/c$, $\vecF_0 = 0$, and
$\calp_0 = (E_0/3)\calI$. Now consider an observer moving at
velocity $\vecv$ relative to the medium. In the observer's frame,
$4\pi B/c$ is the same because the gas temperature $T_0$ is a
world-scalar, and the Lorentz transform to all orders for the energy
gives
\begin{eqnarray}
E & = & \gamma^2 \left(E_0 + 2\frac{\vecv\cdot\vecF_0}{c^2} +
\frac{(\vecv\vecv)}{c^2}\colon\calp_0\right) \\
& = & \gamma^2 \left[1+\frac{1}{3}\left(\frac{v^2}{c^2}\right)\right]
\left(\frac{4 \pi B}{c}\right) \\
& = & \left(\frac{4 \pi B}{c}\right) \left[1 +
\frac{4}{3}\left(\frac{v^2}{c^2}\right) +
O\left(\frac{v^4}{c^4}\right)\right].
\end{eqnarray}
Thus, for this case it is clear that $4\pi B/c - E\sim \beta^2 E$ to
leading order.

Note that using the correct scaling is necessary to obtain sensible
behavior from the equations in the dynamic diffusion limit. If one
assumes that $4\pi B/c - E \sim (\beta/\tau) E$, then in the gas
and radiation energy equations (\ref{gasenergy}) and
(\ref{radenergy}) in the dynamic diffusion limit, the term $\kp
(v^2/c)[(3-R_2)/2] E$ is of higher order than any other term except
perhaps the time derivative. Since this term is non-zero for any
system with non-zero velocity, opacity, and radiation energy density,
this means that there would be no way for the time derivative term to
ever vanish. Thus, a system in the dynamic diffusion limit could never
be in equilibrium unless its velocity or radiation energy were zero
everywhere. Clearly this cannot be correct, since it predicts that our
static, infinite, uniform medium cannot be in equilibrium when seen by
an observer moving by at velocity $\vecv$, even though it is
manifestly in equilibrium in its own rest frame. On the
other hand, if we take $4\pi B/c - E = (4/3) (v^2/c^2) E$, as computed
from the Lorentz transform, it is trivial to verify that equations
(\ref{gasenergy}) and (\ref{radenergy}) correctly give
$\partial(\rho e)/\partial t = \partial E/\partial t = 0$, and $(G^0,
\vecG) = (0, \mathbf{0})$ as well. The observer sees a flux that does work on the gas, but this is precisely canceled by a mismatch between emission and absorption of radiation by the gas, leading to zero net energy transfer.

\end{appendix}

\end{document}